\documentclass[10pt]{article}
\usepackage{amssymb}
\usepackage{amscd}
\usepackage{latexsym}
\usepackage{boldtensors}
\usepackage{amsmath}
\usepackage{amsthm}
\usepackage{graphicx}
\newcommand{\newsec}{\setcounter{equation}{0}\section}%
\newcommand{\Zz}{{\mathbb Z}}
\newcommand{\Nn}{{\mathbb N}}
\newcommand{\Rr}{{\mathbb R}}

\def\be{\begin{equation}}
\def\ee{\end{equation}}
\def\bea{\begin{eqnarray}}
\def\eea{\end{eqnarray}}
\def\Tr{{\rm \,Tr\,}}

\def\d{{\,\rm d}}
\def\i{{\,\rm i}}

\def\0{{\bf 0}}

\def\p{{\bf p}}

\def\P{{\bf P}}

\def\h2m{\frac{\hbar^2}{2m}}
\def\p0{{P_{\beta H^0_N}}}

\def\calZ{{\cal Z}}

\newtheorem*{theorem}{Theorem}
\newtheorem*{lemma}{Lemma}
\newtheorem{definition}{Definition}
\newtheorem{proposition}{Proposition}

\theoremstyle{remark}  \newtheorem{remark}{Remark}

\begin{document}
\title{
\large\bf Fourier formula for quantum partition functions\footnote{Work supported by OTKA Grant No. K128989.}}
\author{Andr\'as S\"ut\H o\\Wigner Research Centre for Physics\\P. O. B. 49, H-1525 Budapest, Hungary\\
E-mail: suto.andras@wigner.hu\\}
\date{}
\maketitle
\thispagestyle{empty}
\begin{abstract}
\noindent
Fourier expansion of the integrand in the path integral formula for the partition function of quantum systems leads to a deterministic expression which, though still quite complex, is easier to process than the original functional integral. It therefore can give access to problems that eluded solution so far. Here we derive the formula;
applications to the problem of Bose-Einstein condensation are presented in the papers  arXiv:2108.02659 [math-ph] and 	arXiv:2208.08931 [math-ph].
\end{abstract}

\newsec{The result}

The Feyman-Kac formula [F1, K1,2] provides a stochastic representation of the partition function of quantum systems [F2, G1-4]. Working on it one can arrive at a deterministic expression as follows.

\begin{theorem}
Consider $N$ identical bosons or fermions on a $d\geq 1$-torus of side $L$ which interact via a pair potential $u:\Rr^d\to\Rr$ of the following properties:
\begin{enumerate}
\item
$u(-x)=u(x)$.
\item
The Fourier-transform $\hat{u}$ of $u$ exists and $\hat{u}\in L^1\cap C(\Rr^d)$.
\item
$u(x)=O\left(|x|^{-d-\eta}\right)$ with some $\eta>0$ as $x\to\infty$.
\end{enumerate}

\noindent
Then the canonical partition function of the system at inverse temperature $\beta$ can be set in the form
\bea\label{QNL}
Q_{N,L}=\exp\left\{-\frac{\beta \hat{u}(0)N(N-1)}{2L^d}\right\} \sum_{p=1}^N\epsilon(p)
\frac{1}{p!}\sum_{n_1,\dots,n_p\geq 1:\sum_1^p n_l=N}\ \prod_{l=1}^p\frac{1}{n_l}
\nonumber\\
\sum_{\alpha^2_1,\alpha^3_1,\alpha^3_2,\dots,\alpha^N_{N-1}=0}^\infty \
\prod_{1\leq j<k\leq N}\frac{\left(-\beta\right)^{\alpha^k_j}}{\alpha^k_j !} \prod_{r=1}^{\alpha^k_j}
\frac{1}{L^d} \sum_{z^k_{j,r}\in\Zz^d\setminus\{0\}}\hat{u}\left(\frac{z^k_{j,r}}{L}\right) \int_0^1\d t^k_{j,r}
\nonumber\\
\left[
\prod_{l=1}^p \delta_{Z^l_1,0} \sum_{z\in\Zz^d} \exp\left\{-\frac{\pi \lambda_\beta^2}{L^2}\sum_{q\in C_l}\int_0^1\left[z+Z_q(t)\right]^2\d t\right\}
\right].
\nonumber\\
\eea
Here $\epsilon(p)=1$ for bosons and $(-1)^{N-p}$ for fermions, and $\lambda_\beta=\sqrt{2\pi\hbar^2\beta/m_0}$ is the thermal wave length for particles of mass $m_0$. There is a summation with respect to $\alpha^k_j$ for every pair $1\leq j<k\leq N$, and the quantity in the square brackets is under all these summations/integrals. Furthermore,
\be
N_l=\sum_{l'=1}^l n_{l'},\quad C_l=\{N_{l-1}+1,N_{l-1}+2,\dots,N_l\}
\ee
and for $q\in C_l$
\bea\label{Zqt}
Z_q(t)=-\sum_{j=1}^{q-1}\sum_{k=q}^{N_l}\sum_{r=1}^{\alpha^{k}_{j}}{\bf 1}\{  t^{k}_{j,r}\geq t\} z^{k}_{j,r} +\sum_{j=q}^{N_l}\sum_{k=N_l+1}^{N}\sum_{r=1}^{\alpha^{k}_{j}}{\bf 1}\{  t^{k}_{j,r}\geq t\} z^{k}_{j,r}\nonumber\\
-\sum_{j=1}^{q}\sum_{k=q+1}^{N_l}\sum_{r=1}^{\alpha^{k}_{j}}{\bf 1}\{  t^{k}_{j,r}<t\} z^{k}_{j,r} +\sum_{j=q+1}^{N_l}\sum_{k=N_l+1}^{N}\sum_{r=1}^{\alpha^{k}_{j}}{\bf 1}\{  t^{k}_{j,r}<t\} z^{k}_{j,r}\ .
\eea
In particular,
\be\label{Z^l_1}
Z^l_1:= Z_{N_{l-1}+1}(0)=-\sum_{j=1}^{N_{l-1}} \sum_{k\in C_l} \sum_{r=1}^{\alpha^{k}_{j}}z^{k}_{j,r}+\sum_{j\in C_l} \sum_{k=N_l+1}^N \sum_{r=1}^{\alpha^{k}_{j}}z^{k}_{j,r}\ .
\ee
\end{theorem}

\vspace{5pt}
\begin{remark}\label{invariance}
It will be seen that $l$ labels a permutation cycle of length $n_l=N_l-N_{l-1}$ ($N_0=0$, $N_p=N$) which stands for a closed effective single-particle trajectory composed of $n_l$ physical particles. The product $\prod_{j<k}[\cdots]$ alone is independent of the order of the $N$ particles. In contrast to this, the definition of $Z_q(t)$ assumes that the $l$th permutation cycle is $(N_{l-1}+1\dots N_l)$. Still, the whole expression does not change if the cycles are taken in a different order, as comparison of Eqs.~(\ref{G}) and (\ref{G-transformed}) in the Lemma below shows it. However, the invariance is deeper in the formulas. Noting that $\hat{u}$ is (also) an even function, this can be seen as follows.
\end{remark}

\begin{proposition}\label{invariance}
For any given set of variables
\[
\left\{\alpha^k_j\geq 1,\ t^k_{j,r}\in[0,1],\ \tilde{z}^k_{j,r}\in\Zz^d\setminus\{0\}\, |\, 1\leq j<k\leq N,\ \ r=1,\dots,\alpha^k_j\right\}
\]
let
\[
{\cal V}=\left\{\{z^k_{j,r}\}\in\left(\Zz^d\setminus\{0\}\right)^{\sum_{j<k}\alpha^k_j} |\, z^k_{j,r}=\pm\tilde{z}^k_{j,r}\right\}.
\]
Then
\[
\sum_{\{z^k_{j,r}\}\in\cal V}\ \prod_{l=1}^p \delta_{Z^l_1,0} \sum_{z\in\Zz^d\setminus\{0\}} \exp\left\{-\frac{\pi \lambda_\beta^2}{L^2}\sum_{q\in C_l}\int_0^1\left[z+Z_q(t)\right]^2\d t\right\}
\]
is independent of the order of the cycles.
\end{proposition}

\noindent
{\em Proof.} It is easier to verify the statement if the particles are numbered by cycles. Particle $q\in C_l$ will carry the double label $li$ where $i=q-N_{l-1}-1$; thus $i\in[0,n_l-1]$. Furthermore, $j, k\in C_l$ will be replaced with the new $j, k$ varying from $0$ to $n_l-1$.
The new notations change $Z_q(t)$ into
\bea\label{Z_{li}}
Z_{li}(t)=&-&\sum_{l'< l}\sum_{j=0}^{n_{l'}-1}\left[\sum_{k=i}^{n_l-1}\sum_{r\,:\,t^{lk}_{l'j,r}\,\geq t}z^{lk}_{l'j,r} +\sum_{k=i+1}^{n_l-1}\sum_{r\,:\,t^{lk}_{l'j,r}< t}z^{lk}_{l'j,r}\right]
\nonumber\\
&+&\sum_{l'> l}\sum_{k=0}^{n_{l'}-1}\left[\sum_{j=i}^{n_l-1}\sum_{r\,:\,t^{l'k}_{lj,r}\,\geq t}z^{l'k}_{lj,r} +\sum_{j=i+1}^{n_l-1}\sum_{r\,:\,t^{l'k}_{lj,r}< t}z^{l'k}_{lj,r}\right]
\nonumber\\
&-&\left[\sum_{j=0}^{i-1}\sum_{k=i}^{n_l-1}\sum_{r\,:\,t^{lk}_{lj,r}\,\geq t}z^{lk}_{lj,r} + \sum_{j=0}^{i}\sum_{k=i+1}^{n_l-1}\sum_{r\,:\,t^{lk}_{lj,r} < t}z^{lk}_{lj,r}\right]
\eea
and
\be
\sum_{q\in C_l}\int_0^1\left[z+Z_q(t)\right]^2\d t= \sum_{i=0}^{n_l-1}\int_0^1\left[z+Z_{li}(t)\right]^2\d t.
\ee
In cycle $l$ the particles above the $i$th are coupled to all the particles in cycles $l'<l$
with a global minus sign and to all the particles in cycles $l'>l$ with a global plus sign.
When the order of the cycles is changed, it is not necessary to change the numbering: all that can happen is that some global signs go wrong. This occurs always in pairs because $z^{lk}_{l'j,r}$ appears in $Z_{l\cdot}(\cdot)$ and $Z_{l'\cdot}(\cdot)$, and only in them, with opposite signs. Replacing the involved variables by their negative restores the correctness of $\{Z_{li}(t)\}$ with a new set of variables within $\cal V$ and the cycles taken in the original order determined by the sequence $(N_l)$. $\quad\Box$

\vspace{5pt}
\begin{remark}
In Eq.~(\ref{QNL}) the sum over $p$-partitions of $N$,
\be\label{partition2}
\frac{1}{p!}\sum_{n_1,\dots,n_p\geq 1:\sum_1^p n_l=N}\ \prod_{l=1}^p\frac{1}{n_l}
=\sum_{n_1\geq\cdots\geq n_p\geq 1:\sum_1^p n_l=N}\prod_{n}\left(\frac{1}{n}\right)^{\mu\left[\{n_l\}_1^p\right]_n}\frac{1}{\mu\left[\{n_l\}_1^p\right]_n!}
\ee
can be replaced by the sum over $p$-compositions of $N$
\bea\label{partition1}
\sum_{(n_1,\dots,n_p)\in\Nn^p:\sum_1^p n_l=N} \frac{1}{N(N-n_1)\cdots(N-n_1-\cdots-n_{p-1})}
\nonumber\\
=\sum_{1\leq N_1<\cdots<N_{p-1}<N}\frac{1}{N(N-N_1)(N-N_2)\cdots(N-N_{p-1})}.
\eea
In (\ref{partition2}) the product on the right side is over the different numbers among $n_1,\dots,n_p$ and $\mu\left[\{n_l\}_1^p\right]_n$ is the multiplicity of $n$ in $\{n_l\}_1^p$.
Both (\ref{partition2}) and (\ref{partition1}) are equal to the fraction of permutations built up from $p$ cycles (for (\ref{partition1}) see later). Thus, summing either of them with respect to $p$ from 1 to $N$ gives 1. In forthcoming applications the following combination of (\ref{partition2}) and(\ref{partition1}) will be utilized: $N$ is partitioned into $p+1$ elements and 1 is decomposed as
\be
1=\frac{1}{N}\left[1+\sum_{n=1}^{N-1}\sum_{p=1}^{N-n}\frac{1}{p!}\sum_{n_1,\dots,n_p\geq 1:\sum_1^p n_l=N-n}\ \frac{1}{\prod_{l=1}^pn_l}\right].
\ee
\end{remark}

\vspace{5pt}
\begin{remark}
In the proof we will obtain
$$\int_0^1\d t^k_{j,\alpha^k_j}\int_0^{t^k_{j,\alpha^k_j}}\d t^k_{j,\alpha^k_j-1} \cdots\int_0^{t^k_{j,2}}\d t^k_{j,1}$$
and replace it by the symmetric
$$\frac{1}{\alpha^k_j!}\int_0^1\d t^k_{j,1}\,\d t^k_{j,2}\cdots \d t^k_{j,\alpha^k_j}\ .$$
This can be done because in the formula (\ref{Zqt}) the times $t^{k}_{j,r}$ are compared to $t$, not to each other, therefore any permutation of $t^{k}_{j,1},\dots,t^{k}_{j,\alpha^k_j}$ rearranges only the sums with respect to $r$.
\end{remark}

\vspace{5pt}
\begin{remark}
The exponents in Eq.~(\ref{QNL}) can be expanded with the result
\bea\label{exprewritten}
\lefteqn{\sum_{z\in\Zz^d} \exp\left\{-\frac{\pi \lambda_\beta^2}{L^2}\sum_{q\in C_l}\int_0^1\left[z+Z_q(t)\right]^2\d t\right\}}
\nonumber\\
&&=\exp\left\{-\frac{\pi n_l \lambda_\beta^2}{L^2}\left[\overline{\left(Z^l_{^\cdot}\right)^2}-\overline{Z^l_{^\cdot}}^2\right]\right\}
\sum_{z\in\Zz^d}\exp\left\{-\frac{\pi n_l \lambda_\beta^2}{L^2}\left(z+\overline{Z^l_{^\cdot}}\right)^2\right\}
\eea
where
\be\label{intZq}
\overline{Z^l_{^\cdot}}=\frac{1}{n_l}\sum_{q\in C_l}\int_0^1 Z_q(t)\d t,\quad \overline{\left(Z^l_{^\cdot}\right)^2}=\frac{1}{n_l}\sum_{q\in C_l}\int_0^1 Z_q(t)^2\d t.
\ee
Performing the time integral and interchanging the order of summations,
\bea\label{intZqbis}
&&n_l\overline{Z^l_{^\cdot}}=\sum_{q\in C_l}\int_0^1 Z_q(t)\d t=-\sum_{j=1}^{N_{l-1}}\sum_{k\in C_l}\sum_{r=1}^{\alpha^{k}_{j}}(k-N_{l-1}-1+t^{k}_{j,r})z^{k}_{j,r}
\nonumber\\
&&-\sum_{\{ j<k\}\subset C_l}(k-j)\sum_{r=1}^{\alpha^{k}_{j}}z^{k}_{j,r}
+\sum_{j\in C_l}\sum_{k=N_l+1}^N \sum_{r=1}^{\alpha^{k}_{j}}(j-N_{l-1}-1+t^{k}_{j,r})z^{k}_{j,r}\nonumber\\
&&\equiv -\sum_{k\in C_l}\sum_{j=1}^{k-1}\sum_{r=1}^{\alpha^{k}_{j}}\left(k-N_{l-1}-1+t^{k}_{j,r}\right)z^{k}_{j,r} +\sum_{j\in C_l}\sum_{k=j+1}^N\sum_{r=1}^{\alpha^{k}_{j}}\left(j-N_{l-1}-1+t^{k}_{j,r}\right)z^{k}_{j,r}.\nonumber\\
\eea
The second form is derived differently, cf. Eq.~(\ref{Zldot}) below, but the identity is evident if we write
\[
k-j=(k-N_{l-1}-1+t^{k}_{j,r})-(j-N_{l-1}-1+t^{k}_{j,r}).
\]
$\overline{\left(Z^l_{^\cdot}\right)^2}$ can be obtained by integration with respect to $t$, but there is still another way described later.
\end{remark}

\vspace{5pt}
\begin{remark}\label{D_l^2=0}
$\overline{\left(Z^l_{^\cdot}\right)^2}-\overline{Z^l_{^\cdot}}^2=0$ if and only if $\overline{\left(Z^l_{^\cdot}\right)^2}=\overline{Z^l_{^\cdot}}^2=0$ and this holds if and only if
\be\label{in-l-cycle}
\alpha^k_j=0 \quad\mbox{whenever}\quad \{j,k\}\cap C_l\neq\emptyset.
\ee
Equation (\ref{in-l-cycle}) implies $Z_q(t)\equiv 0$ and hence $\overline{\left(Z^l_{^\cdot}\right)^2}=\overline{Z^l_{^\cdot}}^2=0$.
For a proof in the other direction one must inspect Eqs.~(\ref{Zqt}), (\ref{Z^l_1}) and (\ref{intZq}). $\overline{\left(Z^l_{^\cdot}\right)^2}-\overline{Z^l_{^\cdot}}^2=0$ if and only  if $Z_q(t)$ is the same constant for all $q\in C_l$ for almost all $t\in [0,1]$. The times $ t^{k}_{j,r}$ occurring in (\ref{Zqt}) are almost surely positive, so there is an interval above 0 in which $Z_{N_{l-1}+1}(t)=Z_{N_{l-1}+1}(0)=Z^l_1$; therefore $Z_q(t)=Z^l_1$ ($=0$ because of $\delta_{Z^l_1,0}$)
for all $q\in C_l$ for almost all $t\in [0,1]$. This however needs the condition (\ref{in-l-cycle}).

A physical interpretation is the following. $\hbar(2\pi/L)Z_q(t)$ may be considered as the shift due to interactions of the momentum of the $q$th particle at "time" $t$ compared to its value in the ideal gas. The shift does not fluctuate if the interaction does not fluctuate, like when the particle is exposed only to a constant external field. This is precisely what (\ref{in-l-cycle}) means, cf. the next remark.
\end{remark}

\vspace{5pt}
\begin{remark}\label{constraint}
In Eq.~(\ref{QNL}) the sum $\sum_{z\in \Zz^d}$ is invariant with respect to the value of $Z_{N_{l-1}+1}(0)$. Imposing $Z^l_1=0$ breaks this invariance. These conditions set a constraint both on the nonzero values of $\alpha^k_j$ and on the number of independent summation variables.
One of $\delta_{Z^l_1,0}$ could be dropped because $\sum_{l=1}^p Z^l_1\equiv 0$. The constraint is absent in the mean-field contribution to the partition function which corresponds to $\alpha^k_j\equiv 0$,
\bea
Q_{N,L}^{\rm m.f.}= \exp\left\{-\frac{\beta\rho\hat{u}(0)(N-1)}{2}\right\}
\sum_{p=1}^N\epsilon(p)
\frac{1}{p!}\sum_{n_1,\dots,n_p\geq 1:\sum_1^p n_l=N}\ \prod_{l=1}^p\frac{1}{n_l}
\sum_{z\in\Zz^d}\exp\left\{-\frac{\pi n_l \lambda_\beta^2}{L^2}z^2\right\}\nonumber\\
\eea
where $\rho=N/L^d$. One can go beyond the  mean-field approximation and still avoid the constraint. $Z^l_1$ depends only on $z^k_{j,r}$ where one and only one of $j$ and $k$ is in
$C_l,$
i.e., in the $l$th cycle. A meaningful completion of the mean-field result is to add terms in which the cycles are decoupled by keeping $\alpha^k_j=0$ if $j$ and $k$ are in different cycles (so their interaction is only via the commonly created mean field) and to choose $\alpha^k_j$ at will if they are in the same cycle. Then, for $q$ in any of the $p$ sets $C_l$
\be
Z_q(t)=-\sum_{j=N_{l-1}+1}^{q-1}\sum_{k=q}^{N_l}\sum_{r=1}^{\alpha^{k}_{j}}{\bf 1}\{  t^{k}_{j,r}\geq t\} z^{k}_{j,r}
-\sum_{j=N_{l-1}+1}^{q}\sum_{k=q+1}^{N_l}\sum_{r=1}^{\alpha^{k}_{j}}{\bf 1}\{  t^{k}_{j,r}<t\} z^{k}_{j,r}
\ee
and in Eq.~(\ref{QNL}) the product over $l=1,\dots,p$ in itself is invariant with respect to the order of the cycles.

To go even further,
\be
A_l:=\sum_{j=1}^{N_{l-1}}\sum_{k\in C_l}\alpha^k_j+\sum_{j\in C_l}\sum_{k=N_l+1}^N\alpha^k_j\geq 2
\ee
is necessary to satisfy $Z^l_1=0$ with nonzero vectors. Such an equation cannot stand alone: if $\alpha^k_j>0$, one of $j$ and $k$ is in another cycle $l'$, implying a coupled equation $Z^{l'}_1=0$. Suppose that $l_1<\cdots<l_s$ ($s\geq 2$) label a maximal set of coupled permutation cycles, giving rise to $s$ homogeneous linear equations. Clearly, $s\leq \frac{1}{2}\sum_{i=1}^s A_{l_i}$, the number of involved pairs $(j,k)$ at $\alpha^k_j$ different times, i.e., the total number of variables in the $s$ equations. Explicitly, these variables are
\[
\bigcup_{1\leq i'<i\leq s}\left\{z^k_{j,r}|j\in C_{l_{i'}},\ k\in C_{l_i},\ r=1,\dots,\alpha^k_j\right\}.
\]
In general, if $j$ is in cycle $l'$ and $k$ is in cycle $l>l'$, $z^k_{j,r}$ appears in two equations with different signs: as $+z^k_{j,r}$ in $Z^{l'}_1=0$ and as $-z^k_{j,r}$ in $Z^{l}_1=0$. For a given set $\{\alpha^k_j\}$ the system of equations may not be soluble with all $z^k_{j,r}$ nonzero. Such an $\{\alpha^k_j\}$ is illicit and is discarded by $\prod_{l=1}^p \delta_{Z^l_1,0}$. On the other hand, if $\{\alpha^k_j\}$ is a valid set and if the number of linearly independent equations is $K_{\{\alpha^k_j\}}$ then $\prod_{l=1}^p \delta_{Z^l_1,0}$ makes $K_{\{\alpha^k_j\}}$ sums over $\Zz^d\setminus\{0\}$ collapse into a single vector leaving a factor $ (L^{-d})^{K_{\{\alpha^k_j\}}}$ uncompensated by sums. This appears as a loss of weight in $Q_{N,L}$ compared to the mean-field and other contributions with uncoupled cycles.

The simplest example of a valid $\{\alpha^k_j\}$ is $s=2$, $\alpha^2_1=2$. With the variables $z^2_{1,1}$ and $z^2_{1,2}$ the two equations are $Z^1_1\equiv z^2_{1,1}+z^2_{1,2}=0$ and $Z^2_1\equiv -z^2_{1,1}-z^2_{1,2}=0$. Thus, $K_{\{\alpha^k_j\}}=1$.

The next simplest example is $s=3$, 1 in cycle $l_1$, 2 in cycle $l_2$, 3 in cycle $l_3$, and $\alpha^2_1=\alpha^3_1=\alpha^3_2=1$.
Then
\[
A_{l_1}=\alpha^2_1+\alpha^3_1=2,\quad A_{l_2}=\alpha^2_1+\alpha^3_2=2,\quad A_{l_3}=\alpha^3_1+\alpha^3_2=2,
\]
and, if $l_1<l_2<l_3$, the three equations are
\[
Z^{l_1}_1\equiv z^2_{1,1}+z^3_{1,1}=0,\quad Z^{l_2}_1\equiv -z^2_{1,1}+z^3_{2,1}=0,\quad Z^{l_3}_1\equiv -z^3_{1,1}-z^3_{2,1}=0.
\]
Choosing any nonzero vector $v$,
\[
z^2_{1,1}=v,\quad z^3_{1,1}=-v,\quad z^3_{2,1}=v
\]
is a solution. Two of the equations represent linearly independent constraints. So the above is the general solution, only one of $z^2_{1,1}, z^3_{1,1}, z^3_{2,1}$ is a free summation variable in Eq.~(\ref{QNL}), $K_{\{\alpha^k_j\}}=2$.

A systematic study of the constraint can be done with the help of graph theory. Consider a graph ${\cal G}_{\{\alpha^k_j\}}$ of $p$ vertices numbered from 1 to $p$ with
$$
b_{l'l}=\sum_{j\in C_{l'}}\sum_{k\in C_l}\alpha^k_j
$$
edges between the vertices $l'$ and $l>l'$. The edge set is valid if to every edge one can assign a nonzero vector so that at any vertex $l$ the sum of the vectors on the incident edges $(l,l_1),(l,l_2),\dots$, taken with minus sign if $l_i<l$ and with plus sign if $l_i>l$, is zero.
A solution to this graph problem is presented in the Appendix.
Here is the result:

\noindent
A set $\{\alpha^k_j\}$ is compatible with the constraint $\prod_{l=1}^p \delta_{Z^l_1,0}$ if and only if every maximal connected subgraph of ${\cal G}_{\{\alpha^k_j\}}$ is either a single vertex or a merger through vertices or along edges of circular graphs of arbitrary length.

A simple argument shows (see Proposition~\ref{K=V-1}) that for a connected merger graph of $V$ vertices the number $K$ of independent equations for the edge variables is $V-1$, whatever be the number of edges. Let the $\sum_{1\leq l'<l\leq p}b_{l'l}$
edges of ${\cal G}_{\{\alpha^k_j\}}$ link the $p$ vertices into $m_{\{\alpha^k_j\}}$ maximal connected subgraphs of $V_1,\dots,V_{m_{\{\alpha^k_j\}}}$ vertices, respectively. The $i$th one contributes to $K_{\{\alpha^k_j\}}$ with $V_i-1$, so
\be
K_{\{\alpha^k_j\}}=\sum_{i=1}^{m_{\{\alpha^k_j\}}}(V_i-1)=p-m_{\{\alpha^k_j\}}.
\ee
The cycles can be coupled in so many different ways that the coupling always overcompensates the loss and results in a decrease of the free energy density. This we illustrate with an example.
In each of its appearances $L^{-d}$ is multiplied with $\beta$ times $\hat{u}$ at different points of the dual space. One can represent the latter by $\hat{u}(0)$ (which is positive for stable pair potentials), so the loss is a dimensionless factor $\sim (\beta\hat{u}(0)/L^{d})^{K_{\{\alpha^k_j\}}}$.
Let $p=cN$ with some $c\leq1$. Suppose that ${\{\alpha^k_j\}}$ is chosen so that we have $aN$ isolated cycles ($a<c$) and the remaining $(c-a)N$ cycles are coupled in 2-circles. Then $m_{\{\alpha^k_j\}}=(a+(c-a)/2)N$, $K_{\{\alpha^k_j\}}=(c-a)N/2$, and the loss is $(\beta\rho\hat{u}(0)/N)^{(c-a)N/2}$. A lower bound to the number of possibilities is
\[
{cN\choose aN} \frac{[(c-a)N]!}{[(c-a)N/2]!\, 2^{(c-a)N/2}}=\frac{(cN)!}{(aN)!\,[(c-a)N/2]!\,2^{(c-a)N/2}}
\]
so the overall factor we obtain by coupling is not smaller than
\[
(\beta\rho\hat{u}(0)/N)^{(c-a)N/2}\frac{(cN)!}{(aN)!\,[(c-a)N/2]!\,2^{(c-a)N/2}} \sim \left[\frac{\beta\rho\hat{u}(0)}{e(c-a)} \right]^{(c-a)N/2}\left(\frac{c^c}{a^a}\right)^N.
\]
Choosing $a$ close enough to $c$, this expression is exponentially increasing with $N$. Its maximum with respect to $a$ may not be the largest exponential factor obtainable from coupling.
The point is that for arbitrarily small $\beta\rho\hat{u}(0)$ the cycles are coupled together in a macroscopic number. Note that the coupling increases also the fluctuation of the shift $Z_q(t)$ of the momenta, hence $\overline{\left(Z^l_{^\cdot}\right)^2}~-\overline{Z^l_{^\cdot}}^2$. This is a source of loss to be taken into account, that decreases the overall factor found above.
\end{remark}

\begin{remark}
Asymptotically, when $N$ and $L$ are large, $Q_{N,L}$ becomes
\bea\label{QN}
\lefteqn{
Q_{N,L}\sim \exp\left\{-\frac{\beta\rho\hat{u}(0)N}{2}\right\}
\sum_{p=1}^N\epsilon(p)
\frac{1}{p!}\sum_{n_1,\dots,n_p\geq 1:\sum_1^p n_l=N}\ \prod_{l=1}^p\frac{1}{n_l}   }
\nonumber\\
&&\sum_{\alpha^2_1,\alpha^3_1,\alpha^3_2,\dots,\alpha^N_{N-1}=0}^\infty
\Delta_{\{\alpha^k_j\},\{n_l\}} \left(L^{-d}\right)^{K_{\{\alpha^k_j\}}}
 \prod_{1\leq j<k\leq N}
\frac{ \left(-\beta\right)^{\alpha^k_j}}{\alpha^k_j !}
\prod_{r=1}^{\alpha^k_j}\int_0^1\d t^k_{j,r} \int \d x^k_{j,r}\ \hat{u}\left(x^k_{j,r}\right)
\nonumber\\
&&  \hspace{2.5cm}
\left[
\delta(X^1_1,\dots,X^p_1) \prod_{l=1}^p \sum_{z\in\Zz^d}
\exp\left\{-\frac{\pi \lambda_\beta^2}{L^2}\sum_{q\in C_l}\int_0^1\left[z+LX_q(t)\right]^2\d t\right\}
\right].
\nonumber\\
\eea
Above
$X^l_1$ and $X_q(t)$ are obtained from $Z^l_1/L$ and $Z_q(t)/L$, respectively, by replacing $z^{k}_{j,r}/L$ with the continuous variable $x^{k}_{j,r}$ for all $j,k,r$. The factor $(L^{-d})^{K_{\{\alpha^k_j\}}}$ appears explicitly because
\bea
\lefteqn{
\left[ \prod_{1\leq j<k\leq N}\prod_{r=1}^{\alpha^k_j}
\frac{1}{L^d} \sum_{z^k_{j,r}\in\Zz^d\setminus\{0\}}\hat{u}\left(\frac{z^k_{j,r}}{L}\right)\right] \prod_{l=1}^p \delta_{Z^l_1,0}
}  \nonumber\\
&&\sim\left(L^{-d}\right)^{K_{\{\alpha^k_j\}}}
\int\prod_{1\leq j<k\leq N}\prod_{r=1}^{\alpha^k_j}\d x^k_{j,r}\ \hat{u}\left(x^k_{j,r}\right) \delta(X^1_1,\dots,X^p_1)
\eea
where $\delta(X^1_1,\dots,X^p_1)$ 
restricts the multiple integral to a $d\left(\sum_{j<k}\alpha^k_j-K_{\{\alpha^k_j\}}\right)$-dimensional manifold on which $X^1_1=\cdots=X^p_1=0$.
Moreover, $\Delta_{\{\alpha^k_j\},\{n_l\}}=1$ if ${\cal G}_{\{\alpha^k_j\}}$ is a merger graph and is zero otherwise; we include this constraint also explicitly.
As in Eqs.~(\ref{exprewritten}) and (\ref{intZq}),
\bea
\lefteqn{
\sum_{z\in\Zz^d}\exp\left\{-\frac{\pi \lambda_\beta^2}{L^2}\sum_{q\in C_l}\int_0^1\left[z+LX_q(t)\right]^2\d t\right\}   }
\nonumber\\
&&=\exp\left\{-\pi n_l \lambda_\beta^2\left[\overline{\left(X^l_{^\cdot}\right)^2}-\overline{X^l_{^\cdot}}^2\right]\right\}
\sum_{z\in\Zz^d}\exp\left\{-\frac{\pi n_l \lambda_\beta^2}{L^2}\left(z+L\overline{X^l_{^\cdot}}\right)^2\right\}
\eea
where
\be
\overline{X^l_{^\cdot}}=\frac{1}{n_l}\sum_{q\in C_l}\int_0^1 X_q(t)\d t,\quad   \overline{\left(X^l_{^\cdot}\right)^2}=\frac{1}{n_l}\sum_{q\in C_l}\int_0^1 X_q(t)^2\d t.
\ee
\end{remark}

\vspace{10pt}\noindent
The heart of the proof of the Theorem is the following lemma.

\begin{lemma}
Let $W^\beta_{xy}(\d\omega)$ denote the Wiener measure on the $d$-torus $\Lambda$ of side $L$ defined on the trajectories that start in $x$ at time 0 and end in $y$ at time $\beta$. Choose $p-1$ integers $1\leq N_1<\cdots<N_{p-1}<N$ and define $n_l=N_l-N_{l-1}$, $l=1,\dots,p$ ($N_0=0$, $N_p=N$). Let moreover
\bea\label{G}
G\left[\{n_l\}_1^p\right]:=\int_\Lambda \d x_1 \int W_{x_1x_1}^{n_1\beta}(\d\omega_1) \prod_{0\leq j<k\leq n_1-1} \exp\left\{-\int_0^\beta u_L(\omega_1(k\beta+t)-\omega_1(j\beta+t)) \d t\right\} \nonumber\\
 \cdots \int_\Lambda \d x_p \int W_{x_px_p}^{n_p\beta}(\d\omega_p) \prod_{0\leq j<k\leq n_p-1} \exp\left\{-\int_0^\beta u_L(\omega_p(k\beta+t)-\omega_p(j\beta+t)) \d t\right\}
\nonumber\\
\prod_{1\leq l'<l\leq p}\prod_{j=0}^{n_{l'}-1}\prod_{k=0}^{n_l-1} \exp\left\{-\int_0^\beta u_L(\omega_l(k\beta+t)-\omega_{l'}(j\beta+t)) \d t\right\}
\eea
where
\be\label{u_L(x)}
u_L(x)=\sum_{z\in\Zz^d} u(x+Lz).
\ee
Then
\bea\label{G-transformed}
\lefteqn{
G\left[\{n_l\}_1^p\right]=\exp\left\{-\frac{\beta \hat{u}(0)N(N-1)}{2L^d}\right\}   }
\nonumber\\
&&\times\sum_{\alpha^2_1,\alpha^3_1,\alpha^3_2,\dots,\alpha^N_{N-1}=0}^\infty\
\prod_{1\leq j<k\leq N} \left(\frac{-\beta}{L^d}\right)^{\alpha^k_j} \sum_{z^k_{j,1},z^k_{j,2},\ldots,z^k_{j,\alpha^k_j}\in\Zz^d\setminus\{0\}}
\hat{u}\left(\frac{z^k_{j,1}}{L}\right) \hat{u}\left(\frac{z^k_{j,2}}{L}\right)\cdots \hat{u}\left(\frac{z^k_{j,\alpha^k_j}}{L}\right)
\nonumber\\
&&
\int_0^1\d t^k_{j,\alpha^k_j}\int_0^{t^k_{j,\alpha^k_j}}\d t^k_{j,\alpha^k_j-1} \cdots\int_0^{t^k_{j,2}}\d t^k_{j,1}\left[ \prod_{l=1}^p \delta_{Z^l_1,0}  \sum_{z\in\Zz^d} \exp\left\{-\frac{\pi \lambda_\beta^2}{L^2}\sum_{q\in C_l}\int_0^1\left[z+Z_q(t)\right]^2\d t\right\}\right].\nonumber\\
\phantom{a}
\eea
\end{lemma}

\newsec{Proof of the Theorem}

We must prove that
\bea\label{Qfinal}
Q_{N,L}&=&\sum_{p=1}^N\frac{\epsilon(p)}{p!} \sum_{n_1,\dots,n_p\geq 1:\sum_1^p n_l=N}\frac{G\left[\{n_l\}_1^p\right]}{\prod_{l=1}^p n_l}
\nonumber\\
&=& \frac{1}{N}\sum_{p=1}^N\epsilon(p)
\sum_{(n_1,\dots,n_p)\in\Nn^p:\sum_1^p n_l=N} \frac{G\left[\{n_l\}_1^p\right]}{(N-n_1)\cdots(N-n_1-\cdots-n_{p-1})}
\nonumber\\
\eea
i.e., that the partition function is an average of $\epsilon(p)G\left[\{n_l\}_1^p\right]$ over the partitions/compositions of $N$. For a system whose Hamiltonian is
\be
H_{N,L}=-\frac{\hbar^2}{2m_0}\sum_{i=1}^N\Delta_i+\sum_{1\leq i<j\leq N}u_L(x_j-x_i)
\ee
the canonical partition function on the torus $\Lambda$ in path integral representation reads
\bea\label{Qoriginal}
Q_{N,L}=\Tr P_\pm e^{-\beta H_{N,L}}=\frac{1}{N!}\sum_{\pi\in S_N}\epsilon_\pm(\pi) \int_{\Lambda^N}\d x_1\cdots\d x_N
\int W^\beta_{x_1x_{\pi(1)}}(\d\omega_1)\cdots \int W^\beta_{x_Nx_{\pi(N)}}(\d\omega_N)
\nonumber\\
 \prod_{1\leq j<k\leq N}\exp\left\{-\int_0^\beta u_L(\omega_k(t)-\omega_j(t))\d t\right\}.
\eea
Here $\pm$ refers to bosons or fermions, respectively, and
\be
P_\pm=\frac{1}{N!}\sum_{\pi\in S_N}\epsilon_\pm(\pi)P_\pi
\ee
is the orthogonal projection to the symmetric or antisymmetric subspace of the $N$-particle Hilbert space: $S_N$ is the symmetric group of $N$ elements, $P_\pi$ is the unitary representation of $\pi$, $\epsilon_+(\pi)\equiv 1$ and $\epsilon_-(\pi)$ is the sign of the permutation $\pi$. Furthermore, $W^\beta_{xy}(\d\omega)$ is the conditional Wiener measure adapted to periodic boundary conditions on a cube $\Lambda$ of side $L$:

\noindent
Let $f$ be a $L\Zz^d$-periodic functional over the space of trajectories $\omega:[0,\beta]\to\Rr^d$ where periodicity is understood in the sense that $f$ depends only on $\left\{\omega(t)_i({\rm mod}\,L)|i=1,\dots,d,\,t\in [0,\beta]\right\}$. Then by definition
\be\label{W}
\int W^\beta_{xy}(\d\omega)f(\omega)=\sum_{z\in\Zz^d}\int P^\beta_{x,y+Lz}(\d\omega)f(\omega),
\ee
where $P^\beta_{xy}$ is the Brownian bridge measure in $\Rr^d$. The measure of the set $\{\omega|\omega(0)=x,\omega(\beta)=y\}$ is
\be\label{norms}
\int P^\beta_{xy}(\d\omega)=\lambda_\beta^{-d}e^{-\pi(x-y)^2/\lambda_\beta^2}=:\psi_\beta(x-y)
\ee
with the contraction property
\be
\int\d r\int P^{\beta_1}_{xr}(\d\omega_1) P^{\beta_2}_{ry}(\d\omega_2)=\int P^{\beta_1+\beta_2}_{xy}(\d\omega).
\ee
Using invariance by space translations of $P^\beta_{xy}$ it is easy to verify that this property extends to $W^\beta_{xy}$,
\be
\int_\Lambda \d r\int W^{\beta_1}_{xr}(\d\omega_1) W^{\beta_2}_{ry}(\d\omega_2)=\int W^{\beta_1+\beta_2}_{xy}(\d\omega).
\ee
In general, if the trajectories $\omega_1$, $\omega_2$ with $\omega_1(0)=x$, $\omega_1(\beta_1)=\omega_2(0)=r$, $\omega_2(\beta_2)=y$ are continuous and $f(\omega_1,\omega_2)$ is a continuous functional then
\be\label{contraction-general}
\int\d r\int P^{\beta_1}_{xr}(\d\omega_1) P^{\beta_2}_{ry}(\d\omega_2)f(\omega_1,\omega_2)= \int P^{\beta_1+\beta_2}_{xy}(\d\omega) \tilde{f}(\omega)
\ee
where $f(\omega_1,\omega_2)=\tilde{f}(\omega_1\circ\omega_2)$ and $\omega_1\circ\omega_2$ is the concatenation of $\omega_1$ and $\omega_2$. This follows from invariance also by time translations,
\be
\int P^\beta_{xy}(\d\omega)\equiv \int P^{0\beta}_{xy}(\d\omega)=\int P^{t,\beta+t}_{xy}(\d\omega)
\ee
via the construction of the Wiener measure as the limit of finite distributions: if
$ 0<t_1<\cdots<t_{n}<\beta_1<\tau_1<\cdots<\tau_{m}<\beta_1+\beta_2,$
$x_i=\omega_1(t_i)$ and $y_j=\omega_2(\tau_j)$, then both sides of Eq.~(\ref{contraction-general}) are equal to
\begin{eqnarray*}
\lefteqn{
\lim_{n,m\to\infty}\int\psi_{t_1}(x_1-x)\psi_{t_2-t_1}(x_2-x_1)\cdots\psi_{\beta_1-t_{n}}(r-x_{n})\psi_{\tau_1-\beta_1}(y_1-r) \psi_{\tau_2-\tau_1}(y_2-y_1)
}
\nonumber\\
&&\cdots\psi_{\beta_1+\beta_2-\tau_{m}}(y-y_{m})
{\bar f}(x,x_1,\dots,x_{n},r,y_1,\dots,y_{m},y) \d x_1\dots\d x_{n}\d r\d y_1\dots\d y_{m}
\end{eqnarray*}
provided that all the time intervals go to zero and ${\bar f}$ is a suitable approximation of and converging to ${\tilde f}$ in the above limit.
If $f$ is also periodic in both $\omega_1$ and $\omega_2$ then
\be\label{contraction-torus}
\int_\Lambda \d r\int W^{\beta_1}_{xr}(\d\omega_1) W^{\beta_2}_{ry}(\d\omega_2)f(\omega_1,\omega_2) =\int W^{\beta_1+\beta_2}_{xy}(\d\omega)\tilde{f}(\omega).
\ee
Here we focused on the particularities of the toroidal geometry; for more details about the general theory see [G1], [LHB].

The permutations form cycles; if one integrates with respect to all but one $x_i$ within each cycle, then by a repeated application of the contraction (\ref{contraction-torus}) the $N$-times integral in Eq.~(\ref{Qoriginal}) transforms into the $p$-times integral in Eq.~(\ref{G}). This one depends on $\pi$ only via $n_1,\dots,n_p$, the lengths of the cycles that constitute the permutation. In fact, the total potential energy is independent of $\pi$ and the sign depends only on the number of cycles: a cycle of length $n$ gives a factor $(-1)^{n-1}$, so the sign of a permutation $\pi$ of $p$ cycles is $\epsilon_-(\pi)=(-1)^{\sum_{l=1}^p (n_l-1)}=(-1)^{N-p}$. Therefore in (\ref{Qoriginal}) each conjugation class can be represented by any of its elements if one inserts a factor that gives the number of elements of the class. This yields the standard form of the partition function shown in the first line of Eq.~(\ref{Qfinal}): $Q_{N,L}$ is the average of $\epsilon(p)G\left[\{n_l\}_1^p\right]$ over the conjugation classes of $S_N$.
The form given in the second line of (\ref{Qfinal}) is slightly more tricky, because each conjugation class that contains cycles of different lengths is decomposed into as many parts as the possible orderings of the cycle lengths, and the associated factors depend on the order of the cycles. The result is an average based on a different resolution of the unity,
\bea\label{resol}
1=\frac{1}{N}\left(1+\sum_{n_1=1}^{N-1}\frac{1}{N-n_1}\left(1+\sum_{n_2=1}^{N-n_1-1}\frac{1}{N-n_1-n_2}
\left(1+\cdots\right)\right)\cdots\right)\nonumber\\ =\frac{1}{N}+\sum_{n_1=1}^{N-1}\frac{1}{N(N-n_1)}+\sum_{n_1=1}^{N-1}\sum_{n_2=1}^{N-n_1-1}\frac{1}{N(N-n_1)(N-n_1-n_2)}+\cdots
\eea
This expression is obtained by "skew" simplification from
\be\label{resol2}
1=\frac{1}{N!}\sum_{n_1=1}^N\frac{(N-1)!}{(N-n_1)!}\sum_{n_2=1}^{N-n_1}\frac{(N-n_1-1)!}{(N-n_1-n_2)!}\cdots.
\ee
To arrive at Eq.~(\ref{resol2}) we write $\pi=\cdots\pi_3\pi_2\pi_1$ where $\pi_1$ is the cycle that contains 1, $\pi_2$ is the cycle that contains the smallest number not in $\pi_1$, $\pi_3$ is the cycle that contains the smallest number not in $\pi_1$ and $\pi_2$, and so on. The number of permutations with 1 in a cycle of length $n_1$ is
$${N-1\choose n_1-1} (n_1-1)!=\frac{(N-1)!}{(N-n_1)!},$$
and the other factors are obtained similarly. The proof of the theorem is done provided the Lemma is verified.

\newsec{Proof of the Lemma}

The main steps of the proof are as follows.

\noindent
(1) Discrete-time approximation: the time integrals of the pair potentials are replaced by their Riemann approximating sum. Then the Wiener measure depends only on a finite sequence of values that the trajectory assumes in the selected instants, and appears as a product of a finite number of (periodized) Gaussians.\\
(2) Each Gaussian and each Boltzmann factor is Fourier-expanded, making possible the integration with respect to the spatial variables. This introduces relations among the Fourier variables and permits to eliminate those associated with the Gaussians, i.e., with the kinetic energy.\\
(3) One takes the limit that restores the continuous time.

\vspace{10pt}
The functional integration could be short-circuited by starting with the Trotter formula for $\exp\{-\beta H_{N,L}\}$ and stopping at the finite-product stage. However, the bulk of the work, points (2) and (3), would not change.

\subsection{Discrete-time approximation}

\noindent
The interval $[0,\beta]$ is divided into $m$ subintervals of length $\beta/m$.  Let
\be\label{xl-km+i}
x^l_{km+i}=\omega_l((km+i)\beta/m),\qquad l=1,\ldots
,p,\quad k=0,\ldots,n_l-1,\quad i=1,\ldots,m,\qquad x^l_{n_l m}\equiv x^l_0=x_l,
\ee
cf. Eq.~(\ref{G}). Thus, the spatial integration variables will be $x^l_1,\dots, x^l_{n_l m}$ for $l=1\dots,p$. Then, with the notation
\be
E_m(x)=\exp\left\{-\frac{\beta}{m} u_L(x)\right\}
\ee
the discrete-time approximation of $G\left[\{n_l\}_1^p\right]$ is
\bea
G_m\left[\{n_l\}_1^p\right]= \int_{\Lambda^{mN}} \prod_{l=1}^p\prod_{j=1}^{n_l m}\d x^l_j\int\prod_{l=1}^p\prod_{j=1}^{n_l m} W^{\beta/m}_{x^l_{j-1}x^l_j}(\d\omega^l_j) \prod_{i=1}^m \prod_{l=1}^p\ \prod_{0\leq j<k\leq n_l-1}E_m(x^l_{km+i}-x^l_{jm+i})\nonumber\\
\prod_{i=1}^m\, \prod_{1\leq l'<l\leq p}\prod_{j=0}^{n_{l'}-1}\prod_{k=0}^{n_{l}-1} E_m(x^l_{km+i}-x^{l'}_{jm+i}).
\eea

\subsection{ Fourier expansions}

Let us start with the Wiener measure. From Eqs.~(\ref{W}) and (\ref{norms})
\be
\int W^{\beta/m}_{xy}(\d\omega)=\sum_{z\in\Zz^d}\int P^{\beta/m}_{x,y+Lz}(\d\omega) =\lambda_{\beta/m}^{-d}\sum_{z\in\Zz^d}e^{-\pi(x-y+Lz)^2/\lambda_{\beta/m}^2} =\frac{1}{L^d}\sum_{z\in\Zz^d}e^{-\pi\lambda_{\beta/m}^2z^2/L^2} e^{\i\frac{2\pi}{L}z\cdot (x-y)}.
\ee
The rightmost form, obtained via the Poisson summation formula, is the Fourier representation we need. Denoting the dual vector of $x^l_{j}- x^l_{j-1}$ by $v^l_{j,j-1}$ ($j=1,\dots,n_lm$),
\bea\label{prodW}
\int\prod_{j=1}^{n_l m} W^{\beta/m}_{x^l_{j-1}x^l_j}(\d\omega^l_j) =\frac{1}{L^{dn_lm}} \sum_{v^l_{10},v^l_{21},\ldots,v^l_{n_lm,n_lm-1}\in\Zz^d} \exp\left\{-\frac{\pi\lambda_{\beta}^2}{mL^2}\sum_{j=1}^{n_lm}(v^l_{j,j-1})^2\right\} \nonumber\\
\times \exp\left\{\i\frac{2\pi}{L}\sum_{j=1}^{n_lm}v^l_{j,j-1}\cdot(x^l_j-x^l_{j-1})\right\}.
\eea
Now
\be
\sum_{j=1}^{n_lm}v^l_{j,j-1}\cdot(x^l_j-x^l_{j-1})=\sum_{j=1}^{n_lm}(v^l_{j,j-1}-v^l_{j+1,j})\cdot x^l_j \quad \mbox{where}\quad  v^l_{n_lm+1,n_lm}\equiv v^l_{10}.
\ee
Introduce
\be
v^l_j=v^l_{j,j-1}-v^l_{j+1,j}\ ,\quad j=1,\ldots,n_lm.
\ee
There are only $n_lm-1$ linearly independent differences,
\be\label{sumvlj}
\sum_{j=1}^{n_lm} v^l_j=0.
\ee
The new variables $\{v^l_j\}_{j=1}^{n_lm-1}$ together with $v^l_{10}$ form a linearly independent set. Utilizing them,
\be
v^l_{j,j-1}=v^l_{10}+\sum_{j'=j}^{n_lm}v^l_{j'},\qquad j=1,\ldots,n_lm,
\ee
\be
\sum_{j=1}^{n_lm}(v^l_{j,j-1})^2= \sum_{j=1}^{n_1m}(v^l_{10}+\sum_{j'=j}^{n_lm}v^l_{j'})^2,
\ee
\be
\sum_{j=1}^{n_lm}v^l_{j,j-1}\cdot(x^l_j-x^l_{j-1})=\sum_{j=1}^{n_lm}v^l_j\cdot x^l_j.
\ee
Substituting the last two expressions into Eq.~(\ref{prodW}) and replacing the summation variables $v^l_{j+1,j}$ by $v^l_j$ for $j=1,\dots,n_lm-1$ while keeping $v^l_{10}$,
\bea\label{Wxj-1xj}
\int\prod_{j=1}^{n_l m} W^{\beta/m}_{x^l_{j-1}x^l_j}(\d\omega^l_j) =\frac{1}{L^{dn_lm}}
\sum_{v^l_{10}\in\Zz^d}\ \sum_{v^l_{1},v^l_{2},\ldots,v^l_{n_lm-1}\in\Zz^d} \exp\left\{-\frac{\pi\lambda_{\beta}^2}{mL^2}\sum_{j=1}^{n_lm}(v^l_{10}+\sum_{j'=j}^{n_lm}v^l_{j'})^2\right\}
\nonumber\\
\prod_{j=1}^{n_l m}\exp\left\{\i\frac{2\pi}{L}v^l_{j}\cdot x^l_{j}\right\}.
\eea
Observe that $v^l_{n_lm}$ could be dropped from the sums over $j'$: adding it to $v^l_{10}$ only shifts $\Zz^d$ by a lattice vector in the summation over $v^l_{10}$.
Because of $v^l_{n_lm}=-\sum_{j=1}^{n_lm-1} v^l_{j}$,
\be
\sum_{j'=j}^{n_lm}v^l_{j'}=-\sum_{j'=1}^{j-1}v^l_{j'}.
\ee

The Fourier expansion of the Boltzmann factors is
\be\label{Ell'}
E_m(x^l_{km+i}-x^{l'}_{jm+i})=\sum_{z_{l'j}^{lk}(i)\in\Zz^d}\hat{E}_m\left(z_{l'j}^{lk}(i)\right) \exp\left\{\i\frac{2\pi}{L}z_{l'j}^{lk}(i)\cdot (x^l_{km+i}-x^{l'}_{jm+i}) \right\}.
\ee
Here
\be\label{hatEmz}
\hat{E}_m(z)=\frac{1}{L^d}\int_\Lambda \exp\left\{-{\mathrm i}\frac{2\pi}{L}z\cdot x\right\}\exp\left\{-\frac{\beta}{m}u_L(x)\right\} \d x =\delta_{z,0} -\frac{\beta}{mL^d}\ \hat{u}(z/L)+\frac{1}{L^d}O(1/m^2)
\ee
with
\be
\hat{u}(z/L)=\int \exp\left\{-\i\frac{2\pi}{L}z\cdot x\right\} u(x) \d x.
\ee
In Eq.~(\ref{hatEmz}) after the Taylor-expansion of $\exp\{-(\beta/m)u_L(x)\}$ we substituted (\ref{u_L(x)}). The summation over $\Zz^d$ combined with the integral over $\Lambda$ is equivalent to integrating over the whole space.
Notice that $\hat{E}_m(z)=\hat{E}_m(-z)=\hat{E}_m(z)^*$. This holds true because $\Lambda$ is a period cell for $\exp\{-{\mathrm i}(2\pi/L)z\cdot x\}\exp\{-(\beta/m)u_L(x)\}$ and, thus, can be chosen to be the cube $[-L/2,L/2)^d$. Furthermore, because $u(x)=u(-x)$, $u_L(x)=u_L(-x)$ as well, cf. Eq.~(\ref{u_L(x)}), implying that $\hat{E}_m(z)$ is indeed a real even function of $z$. Similarly, $\hat{u}(z/L)$ and also the remainder are real even functions of $z$. Some short notations will be useful:
\be
\calZ_{l}=\left(\Zz^d\right)^{mn_l(n_l-1)/2}, \quad
\calZ_{<l}=\left(\Zz^d\right)^{m\sum_{l'<l}n_{l'}n_l}, \quad
\calZ=\left(\Zz^d\right)^{mN(N-1)/2}.
\ee
The elements of these sets are considered as sets of vectors from $\Zz^d$, and the notation
\be\label{E^Z}
\hat{E}_m^Z=\prod_{z\in Z}\hat{E}_m(z)
\ee
will be used for $Z\in\calZ_{l}$, $\calZ_{<l}$, $\calZ$.
Now
\be
\prod_{i=1}^m\ \prod_{0\leq j<k\leq n_l-1} E_m(x^l_{km+i}-x^l_{jm+i}) =\sum_{Z\in\calZ_{l}}\hat{E}_m^Z
\prod_{i=1}^m \exp\left\{\i\frac{2\pi}{L}\sum_{k=1}^{n_l-1}\sum_{j=0}^{k-1} z^{lk}_{lj}(i)\cdot(x^l_{km+i}-x^l_{jm+i})\right\}
\ee
which, after substituting
\be
\sum_{k=1}^{n_l-1}\sum_{j=0}^{k-1} z^{lk}_{lj}(i)\cdot(x^l_{km+i}-x^l_{jm+i})=\sum_{k=0}^{n_l-1} x^l_{km+i}\cdot  \left(\sum_{j=0}^{k-1}z^{lk}_{lj}(i) - \sum_{j=k+1}^{n_l-1}z^{lj}_{lk}(i)\right)
\ee
results in
\be\label{Eml}
\prod_{i=1}^m\ \prod_{0\leq j<k\leq n_l-1} E_m(x^l_{km+i}-x^l_{jm+i}) =\sum_{Z\in\calZ_{l}}\hat{E}_m^Z\prod_{i=1}^m \prod_{k=0}^{n_l-1} \exp\left\{\i\frac{2\pi}{L} x^l_{km+i}\cdot  \left(\sum_{j=0}^{k-1}z^{lk}_{lj}(i) - \sum_{j=k+1}^{n_l-1}z^{lj}_{lk}(i)\right)\right\}.
\ee
Furthermore,
\be
\sum_{1\leq l'<l\leq p}\sum_{k=0}^{n_l-1}\sum_{j=0}^{n_{l'}-1}z^{lk}_{l'j}(i)\cdot \left(x^l_{km+i}-x^{l'}_{jm+i}\right) =\sum_{l=1}^p\sum_{k=0}^{n_l-1}x^l_{km+i}\cdot\left(\sum_{l'=1}^{l-1}\sum_{j=0}^{n_{l'}-1}z^{lk}_{l'j}(i) -\sum_{l'=l+1}^p\sum_{j=0}^{n_{l'}-1}z^{l'j}_{lk}(i)\right),
\ee
so
\bea\label{Eml'ljk}
\lefteqn{\prod_{i=1}^m\, \prod_{1\leq l'<l\leq p}\prod_{j=0}^{n_{l'}-1}\prod_{k=0}^{n_{l}-1} E_m(x^l_{km+i}-x^{l'}_{jm+i})}
\nonumber\\
&&=\sum_{Z\in\calZ_{<l}}\hat{E}_m^Z\prod_{i=1}^m\prod_{l=1}^p\prod_{k=0}^{n_l-1} \exp\left\{\i\frac{2\pi}{L} x^l_{km+i}\cdot \left(\sum_{l'=1}^{l-1}\sum_{j=0}^{n_{l'}-1}z^{lk}_{l'j}(i) -\sum_{l'=l+1}^p\sum_{j=0}^{n_{l'}-1}z^{l'j}_{lk}(i)\right)\right\}.
\eea
From Eqs.~(\ref{Wxj-1xj}), (\ref{Eml}) and (\ref{Eml'ljk}) one can collect the multiplier of $x^l_{km+i}$.
If we write $G_m\left[\{n_l\}_1^p\right]$ as
\be
G_m\left[\{n_l\}_1^p\right]=\int_{\Lambda^{mN}} \prod_{l=1}^p\prod_{j=1}^{n_lm}\d x^l_j\  G\left[\left\{(x^l_j)_{j=1}^{n_lm},n_l\right\}_{l=1}^p\right]
\ee
then
\bea\label{Gxlj-nl}
\lefteqn{
G\left[\left\{(x^l_j)_{j=1}^{n_lm},n_l\right\}_{l=1}^p\right]=\sum_{Z\in\calZ}\hat{E}_m^Z}
\nonumber\\
&&\prod_{l=1}^p
\left\{\sum_{v^l_{10}\in\Zz^d}\ \sum_{v^l_{1},v^l_{2},\ldots,v^l_{n_lm-1}\in\Zz^d} \left[e^{-\frac{\pi\lambda_{\beta}^2}{mL^2}\sum_{j=1}^{n_lm}(v^l_{10}+\sum_{j'=j}^{n_lm}v^l_{j'})^2}
\prod_{k=0}^{n_l-1}\prod_{i=1}^m\frac{1}{L^d} e^{\i\frac{2\pi}{L}x^l_{km+i}\cdot(v^l_{km+i}- z^l_{km+i})}\right]\right\}
\nonumber\\
\eea
where
\be\label{zlkm+i}
z^l_{km+i}=-\sum_{j=0}^{k-1}z^{lk}_{lj}(i)+\sum_{j=k+1}^{n_l-1}z^{lj}_{lk}(i)
-\sum_{l'=1}^{l-1}\sum_{j=0}^{n_{l'}-1}z^{lk}_{l'j}(i) +\sum_{l'=l+1}^p\sum_{j=0}^{n_{l'}-1}z^{l'j}_{lk}(i).
\ee
The multiple sum (\ref{Gxlj-nl}) is absolutely convergent and can be integrated term by term. In each term there is factorization according to the spatial variables $x^l_{km+i}$, so the integration of the complex units can be made one by one.
Integration with respect to $x^l_j$ in $\Lambda$ makes $v^l_j$ coincide with $z^l_j$ for all $j$
and turns Eq.~(\ref{sumvlj}) into
\be\label{sum-zlj}
\sum_{j=1}^{n_lm}z^l_j=0.
\ee
Thus,
\be\label{Gm1}
G_m\left[\{n_l\}_1^p\right]=\sum_{Z\in\calZ} \left(\prod_{l=1}^p \delta_{\sum_{j=1}^{n_lm}z^l_j,0}\right)\hat{E}_m^Z \sum_{v_l\in\Zz^d} \exp\left\{-\frac{\pi\lambda_{\beta}^2}{mL^2}\sum_{j=1}^{n_lm}(v_l+\sum_{j'=j}^{n_lm}z^l_{j'})^2\right\}.
\ee
Finally, defining
\be
Z^l_j=\sum_{j'=j}^{n_lm}z^l_{j'},\qquad \overline{Z^l_{^\cdot}}=\frac{1}{n_lm}\sum_{j=1}^{n_lm}Z^l_j,\qquad \overline{\left(Z^l_{^\cdot}\right)^2}=\frac{1}{n_lm}\sum_{j=1}^{n_lm}(Z^l_j)^2,
\ee
where we use the same notations for the averages as for their $m\to\infty$ limit in (\ref{exprewritten}), Eq.~(\ref{Gm1}) becomes
\bea\label{Gm2}
\lefteqn{
G_m\left[\{n_l\}_1^p\right] =  \sum_{Z\in\calZ} \left(\prod_{l=1}^p \delta_{Z^l_1,0}\right) \hat{E}_m^Z \sum_{z\in\Zz^d} \exp\left\{-\frac{\pi\lambda_{\beta}^2}{mL^2}\sum_{j=1}^{n_lm}(z+Z^l_j)^2\right\}
} \nonumber\\
&&=\sum_{Z\in\calZ} \left(\prod_{l=1}^p \delta_{Z^l_1,0}\right)\hat{E}_m^Z \exp\left\{-\frac{\pi n_l\lambda_{\beta}^2}{L^2} \left[\overline{\left(Z^l_{^\cdot}\right)^2}-\overline{Z^l_{^\cdot}}^2\right]\right\}\sum_{z\in\Zz^d} \exp\left\{-\frac{\pi n_l\lambda_\beta^2}{L^2}\left(z+\overline{Z^l_{^\cdot}}\right)^2\right\}.
\eea
A consequence of $Z^l_1=0$ is that $\overline{\left(Z^l_{^\cdot}\right)^2}-\overline{Z^l_{^\cdot}}^2=0$ if and only if $Z^l_j=0$ for $j=1,\ldots,n_lm$ which holds if and only if $z^l_j=0$ for $j=1,\ldots,n_lm$. (Note: $Z^l_j=-\sum_{j'=1}^{j-1}z^l_{j'}$ as well, cf. Eq.~(\ref{sum-zlj}).) This is the finite-$m$ equivalent of Remark~\ref{D_l^2=0}.
Two identities in the individual $z$ vectors are noteworthy because they hold for every $i$:
\be
\sum_{k=0}^{n_l-1}z^l_{km+i}
=-\sum_{l'=1}^{l-1}\sum_{j=0}^{n_{l'}-1}\sum_{k=0}^{n_l-1}z^{lk}_{l'j}(i) +\sum_{l'=l+1}^p\sum_{j=0}^{n_{l'}-1}\sum_{k=0}^{n_l-1}z^{l'j}_{lk}(i),
\ee
i.e. the intra-cycle contribution identically vanishes, and
\be
\sum_{l=1}^p \sum_{k=0}^{n_l-1}z^l_{km+i}=0.
\ee
Summing this latter with respect to $i$ yields $\sum_{l=1}^p Z^l_1\equiv 0$.

At this point it is useful to change the notation and number the particles continuously from 1 to $N$. Particle $lk$, the $k$th particle (starting from 0) of the $l$th cycle will carry the number $N_{l-1}+k+1$ when counted continuously, so the identities new$\equiv$old are
\be\label{z-continuous-number}
z^{N_{l-1}+k+1}_{N_{l'-1}+j+1}(i)\equiv z^{lk}_{l'j}(i)\ \mbox{or}\ z^q_{j'}(i)\equiv z^{l,q-N_{l-1}-1}_{l',j'-N_{l'-1}-1}(i),
\quad z_{N_{l-1}+k+1}(i)\equiv z^l_{km+i}\ \mbox{or}\ z_q(i)\equiv z^l_{(q-N_{l-1}-1)m+i}
\ee
for $q\in C_l$ and $j'\in C_{l'}$.
The new notation is better suited to $\hat{E}_m^Z$ which is independent of the cycle structure. Also, the expression (\ref{zlkm+i}) is replaced by the cycle-independent
\be\label{zji}
z_q(i)=-\sum_{j=1}^{q-1} z^q_{j}(i) + \sum_{k=q+1}^N z^{k}_q(i), \qquad q=1,\ldots,N, \quad i=1,\ldots,m,
\ee
and the expanded form of $Z^l_j$, $\overline{Z^l_{^\cdot}}$ and $\overline{\left(Z^l_{^\cdot}\right)^2}$ in terms of the individual $z$ vectors becomes more transparent. For $q\in C_l$ one finds
\bea\label{Zlqi}
\lefteqn{Z^l_{(q-N_{l-1}-1)m+i}=\sum_{i'=i}^m\sum_{k=q}^{N_l}z_{k}(i')+\sum_{i'=1}^{i-1}\sum_{k=q+1}^{N_l}z_{k}(i')}
\\
&&=\sum_{i'=i}^m\left[-\sum_{j=1}^{q-1}\sum_{k=q}^{N_l}z^{k}_{j}(i') +\sum_{j=q}^{N_l}\sum_{k=N_l+1}^Nz^{k}_{j}(i')\right]
+\sum_{i'=1}^{i-1}\left[-\sum_{j=1}^{q}\sum_{k=q+1}^{N_l}z^{k}_{j}(i') +\sum_{j=q+1}^{N_l}\sum_{k=N_l+1}^N z^{k}_{j}(i')\right].
\nonumber
\eea
In particular,
\be\label{Zl1new}
Z^l_1=\sum_{i=1}^m\left[-\sum_{j=1}^{N_{l-1}}\sum_{k\in C_l}z^{k}_{j}(i) +\sum_{j\in C_l}\sum_{k=N_l+1}^N z^{k}_{j}(i)\right].
\ee
Moreover,
\bea\label{Zldot}
\lefteqn{\overline{Z^l_{^\cdot}}=\frac{1}{n_l}\sum_{q\in C_l}\sum_{i=1}^m \left(q-N_{l-1}-1+\frac{i}{m}\right) z_q(i)}\nonumber\\
&&=\frac{1}{n_l}\sum_{q\in C_l}\left[-\sum_{j=1}^{q-1}\sum_{i=1}^m\left(q-N_{l-1}-1+\frac{i}{m}\right)\,z^{q}_{j}(i) +\sum_{k=q+1}^{N}\sum_{i=1}^m\left(q-N_{l-1}-1+\frac{i}{m}\right)\,z^{k}_{q}(i)\right]
\nonumber\\
\eea
and
\bea\label{Zldotsquare}
\overline{\left(Z^l_{^\cdot}\right)^2}&=&\frac{1}{n_l}\sum_{k,k'=0}^{n_l-1}\ \sum_{i,i'=1}^m \min\left\{k+\frac{i}{m},k'+\frac{i'}{m}\right\} z^l_{km+i}\cdot z^l_{k'm+i'}
\nonumber\\
&=&\frac{1}{n_l}\sum_{q,q'\in C_l}\ \sum_{i,i'=1}^m\left(\min\left\{q+\frac{i}{m},q'+\frac{i'}{m}\right\}-N_{l-1}-1\right) z_{q}(i)\cdot z_{q'}(i').
\eea
The last formula becomes complete after (\ref{zji}) is substituted in it.

\subsection{The limit of continuous time}
The result of the limit when $m$ tends to infinity can be seen on the simplest example, that of $N=2$.
Because in this case there is a single pair, the notation can be simplified by writing $z(i)$ instead of $z^2_1(i)$.
Now $\calZ=(\Zz^d)^m$ whose elements are $Z=\{z(1),\ldots,z(m)\}$. There are two partitions of 2: $p=1$, $n_1=2$ and $p=2$, $n_1=n_2=1$.

\vspace{10pt}
\noindent
{\bf One two-particle trajectory}

\vspace{10pt}
\noindent
When $p=1$ then $l=1$, $N_{l-1}=0$, $N=N_1=n_1=2$, so from (\ref{zji}) and (\ref{Zl1new})
\be
z_1(i)=z(i),\quad z_2(i)=-z(i), \quad Z^1_1\equiv 0.
\ee
Moreover,
\be
\overline{Z^1_\cdot}=\frac{1}{2}\sum_{i=1}^m\left[ -\left(1+\frac{i}{m}\right)z(i) +\frac{i}{m}z(i)\right] = - \frac{1}{2}\sum_{i=1}^m z(i),
\ee
\bea
\overline{\left(Z^1_\cdot\right)^2}=\frac{1}{2}\sum_{j,j'=1}^2\sum_{i,i'=1}^m \left(\min\left\{j+\frac{i}{m},j'+\frac{i'}{m}\right\}-1\right) z_j(i)\cdot z_{j'}(i') \nonumber\\
=\frac{1}{2}\left(\sum_{i=1}^m z(i)\right)^2 - \frac{1}{2}\sum_{i,i'=1}^m\frac{|i-i'|}{m} z(i)\cdot z(i')\nonumber\\
= 2 \overline{Z^1_\cdot}^2  - \frac{1}{2}\sum_{i,i'=1}^m\frac{|i-i'|}{m} z(i)\cdot z(i').
\eea
Thus,
\be
\overline{\left(Z^1_\cdot\right)^2}- \overline{Z^1_\cdot}^2= \sum_{i,j=1}^m \left(\frac{1}{4}-\frac{|i-j|}{2m}\right)z(i)\cdot z(j).
\ee

Although it will not be used, we mention that the spectral problem associated with the above quadratic form can be solved exactly.

\begin{proposition}\label{spec(A)}
Let $A$ be the $m\times m$ matrix of elements $A_{ij}=\frac{m}{4}-\frac{|i-j|}{2}$. Then
\[
A^{-1}=
\begin{bmatrix}
2&-1&0&\dots&0&0&1\\
-1&2&-1&\dots&0&0&0\\
\hdotsfor[1.8]{7}\\
0&0&0&\dots&-1&2&-1\\
1&0&0&\dots&0&-1&2
\end{bmatrix},
\]
i.e., minus the discrete Laplace operator with antiperiodic boundary condition $v(j+m)=-v(j)$. The eigenvalues and eigenvectors of $A^{-1}$ are
\[\lambda_{2q-1}=\lambda_{2q}=2\left(1-\cos\frac{(2q-1)\pi}{m}\right)\]
\[v_{2q-1}(j)=\sin\frac{(2q-1)\pi}{m}j,\quad v_{2q}(j)=\cos\frac{(2q-1)\pi}{m}j\qquad (q=1,\dots,\lfloor m/2\rfloor)\]
and
\[\lambda_m=4,\quad v_m(j)=(-1)^j\qquad \mbox{if $m$ is odd}.\]
\end{proposition}
The proof was obtained by making and verifying an ansatz based on numerical calculations of the eigenvalues and eigenvectors of $A$ up to $m=50$.

\vspace{10pt}
\noindent
{\bf Two one-particle trajectories}

\vspace{10pt}
\noindent
In this case
\[
Z^1_1=-Z^2_1=\sum_{i=1}^m z(i)
\]
\[
\overline{Z^1_\cdot}=-\overline{Z^2_\cdot}=\sum_{i=1}^m\frac{i}{m} z(i)
\]
\[
\overline{(Z^1_\cdot)^2}=\overline{(Z^2_\cdot)^2}=\sum_{i,j=1}^m  \min\left\{\frac{i}{m},\frac{j}{m}\right\}z(i)\cdot z(j).
\]

If $z(i)\neq 0$ precisely for $\alpha$ values of $i$, say, for $i_1<i_2<\cdots<i_\alpha$ then everywhere one can replace $\sum_{i=1}^m$ by $\sum_{r=1}^\alpha$ and $i$ in the summand by $i_r$. Because $z(i_r)$ is a summation variable over $\Zz^d\setminus\{0\}$, for its labelling the value of $i_r$ is unimportant and the notation $z_r$ can be used for it. All this permits to rewrite
\be
G_m\left[\{n_l\}_1^p\right]=\sum_{Z\in(\Zz^d)^m}  \left(\prod_{l=1}^p \delta_{Z^l_1,0}\right)\hat{E}_m^Z\ f_{\left[\{n_l\}_1^p\right]}\left(\{i/m\}_{i=1}^m,Z\right)
\ee
where
\be
f_{[2]}\left(\{i/m\}_{i=1}^m,Z\right)=\exp\left\{-\frac{\pi\lambda_\beta^2}{L^2}\sum_{i,j=1}^m\left(\frac{1}{2}-\frac{|i-j|}{m}\right)z(i)\cdot z(j)\right\}
\sum_{z\in\Zz^d}\exp\left\{-\frac{2\pi\lambda_\beta^2}{L^2}\left(z-\frac{1}{2}\sum_{i=1}^m z(i)\right)^2\right\}
\ee
and
\bea
f_{[1,1]}\left(\{i/m\}_{i=1}^m,Z\right)
=
\exp\left\{-\frac{2\pi\lambda_\beta^2}{L^2} \sum_{i,j=1}^m\left[\min\left\{\frac{i}{m},\frac{j}{m}\right\}-\frac{ij}{m^2}\right]z(i)\cdot z(j)\right\}\nonumber\\
\left[\sum_{z\in\Zz^d}\exp\left\{-\frac{\pi\lambda_\beta^2}{L^2}\left(z+\sum_{i=1}^m \frac{i}{m}z(i)\right)^2\right\}\right]^2
\eea
as
\bea
G_m\left[\{n_l\}_1^p\right]
&=&
\sum_{z(1),\dots,z(m)\in\Zz^d}\left(\prod_{l=1}^p \delta_{Z^l_1,0}\right) \hat{E}_m(z(1))\cdots\hat{E}_m(z(m))
f_{\left[\{n_l\}_1^p\right]}\left(\{i/m\}_{i=1}^m,Z\right)
\nonumber\\
&=&
\sum_{\alpha=0}^m \hat{E}_m(0)^{m-\alpha} \sum_{z_1,\dots,z_\alpha\in\Zz^d\setminus\{0\}}\left(\prod_{l=1}^p \delta_{Z^l_1,0}\right) \hat{E}_m(z_1) \cdots \hat{E}_m(z_\alpha)
\nonumber\\
&\times&
\sum_{1\leq i_1<\cdots <i_\alpha\leq m}\,f_{\left[\{n_l\}_1^p\right]}\left(\{i_r/m\}_{r=1}^\alpha,\{z_r\}_{r=1}^\alpha\right).\nonumber\\
\eea
Now
\bea
\sum_{1\leq i_1<\cdots <i_\alpha\leq m}\, f_{\left[\{n_l\}_1^p\right]}\left(\{i_r/m\}_{r=1}^\alpha,\{z_r\}_{r=1}^\alpha\right) =\sum_{i_\alpha=\alpha}^m\ \sum_{i_{\alpha-1}=\alpha-1}^{i_\alpha-1}\cdots \sum_{i_1=1}^{i_2-1}f_{\left[\{n_l\}_1^p\right]}\left(\{i_r/m\}_{r=1}^\alpha,\{z_r\}_{r=1}^\alpha\right)\nonumber\\
=m^\alpha\, \frac{1}{m}\sum_{i_\alpha=\alpha}^m\ \frac{1}{m}\sum_{i_{\alpha-1}=\alpha-1}^{i_\alpha-1}\cdots \frac{1}{m}\sum_{i_1=1}^{i_2-1}f_{\left[\{n_l\}_1^p\right]}\left(\{i_r/m\}_{r=1}^\alpha,\{z_r\}_{r=1}^\alpha\right)\nonumber\\
=m^\alpha \frac{1}{m}\sum_{t_\alpha=\alpha/m}^1\ \frac{1}{m}\sum_{t_{\alpha-1}=(\alpha-1)/m}^{t_\alpha-1/m}\cdots \frac{1}{m}\sum_{t_1=1/m}^{t_2-1/m}f_{\left[\{n_l\}_1^p\right]}\left(\{t_r\}_{r=1}^\alpha,\{z_r\}_{r=1}^\alpha\right)
\nonumber\\
\eea
where each $t_r=i_r/m$ varies by steps $1/m$. From Eq.~(\ref{hatEmz}) we substitute $\hat{E}_m(z)$, dropping the $O(1/m^2)$ term which disappears in the $m\to\infty$ limit. This gives
\bea
\lefteqn{
G_m\left[\{n_l\}_1^p\right]=\sum_{\alpha=0}^m \left(1-\frac{\beta \hat{u}(0)}{mL^d}\right)^{m-\alpha}\left(\frac{-\beta}{L^d}\right)^\alpha  \sum_{z_1,\dots,z_\alpha\in\Zz^d\setminus\{0\}}\left(\prod_{l=1}^p \delta_{Z^l_1,0}\right)\hat{u}(z_1/L) \cdots \hat{u}(z_\alpha/L)
}\nonumber\\
&&\times\frac{1}{m}\sum_{t_\alpha=\alpha/m}^1\ \frac{1}{m}\sum_{t_{\alpha-1}=(\alpha-1)/m}^{t_\alpha-1/m}\cdots \frac{1}{m}\sum_{t_1=1/m}^{t_2-1/m}f_{\left[\{n_l\}_1^p\right]}\left(\{t_r\}_{r=1}^\alpha,\{z_r\}_{r=1}^\alpha\right) =\sum_{\alpha=0}^M \cdots + \sum_{\alpha=M+1}^m \cdots
\nonumber\\
\eea
Taking the modulus of the second sum, $|f_{\left[\{n_l\}_1^p\right]}|$ has the $\alpha$-independent  upper bound
\be
|f_{[2]}|\leq \sum_{z\in\Zz^d} \exp\left\{-\frac{2\pi\lambda_\beta^2}{L^2}z^2\right\},\quad f_{[1,1]}\leq\left[\sum_{z\in\Zz^d} \exp\left\{-\frac{\pi\lambda_\beta^2}{L^2}z^2\right\}\right]^2,
\ee
and the remaining part of the summand is bounded above by
\[
\frac{1}{\alpha !}\left(\frac{\beta \sum_{z\neq 0}|\hat{u}(z/L)|}{L^d}\right)^\alpha.
\]
Therefore
$\lim_{m\to\infty}\sum_{\alpha=M+1}^m\cdots$ is absolutely convergent and goes to zero as $M$ goes to infinity, implying
\bea
G\left[\{n_l\}_1^p\right]
=e^{-\beta\hat{u}(0)/L^d}
\sum_{\alpha=0}^\infty \left(\frac{-\beta}{L^d}\right)^\alpha  \sum_{z_1,\dots,z_\alpha\in\Zz^d\setminus\{0\}}\left(\prod_{l=1}^p \delta_{Z^l_1,0}\right)\hat{u}(z_1/L) \cdots \hat{u}(z_\alpha/L)
\nonumber\\
\times\int_0^1\d t_\alpha\int_0^{t_\alpha}\d t_{\alpha-1}\cdots\int_0^{t_2}\d t_1 f_{\left[\{n_l\}_1^p\right]}\left(\{t_r\}_{r=1}^\alpha,\{z_r\}_{r=1}^\alpha\right).
\nonumber\\
\eea
This ends the proof of the Lemma for $N=2$. Note that for $p=2$ the constraint $\sum_{r=1}^\alpha z_r=0$ acts both on $\alpha$ and on the set of variables $z_1,\dots,z_\alpha$: for $\alpha=1$ the sum over $\Zz^d\setminus \{0\}$ is empty and for $\alpha>1$ only $\alpha-1$ variables can be chosen freely from $\Zz^d\setminus\{0\}$, meaning that these terms are of order $L^{-d}$.

For a general $N$, consider first the limit of the exponents in Eq.~(\ref{Gm2}). In the first line
\be
\frac{1}{m}\sum_{j=1}^{n_lm}(z+Z^l_j)^2=\sum_{k=0}^{n_l-1}\frac{1}{m}\sum_{i=1}^m (z+Z^l_{km+i})^2 =\sum_{q\in C_l}\frac{1}{m}\sum_{i=1}^m (z+Z^l_{(q-N_{l-1}-1)m+i})^2.
\ee
Substituting for $Z^l_{(q-N_{l-1}-1)m+i}$ the expression (\ref{Zlqi}) and keeping only the nonzero $z$ vectors,
\be
\sum_{i'=i}^m z^{k}_{j}(i')=\sum_{r=1}^{\alpha^{k}_{j}}{\bf 1}\left\{\frac{i^{k}_{j,r}}{m}\geq \frac{i}{m}\right\}z^{k}_{j}\left(i^{k}_{j,r}\right), \qquad \sum_{i'=1}^{i-1} z^{k}_{j}(i')=\sum_{r=1}^{\alpha^{k}_{j}}{\bf 1}\left\{\frac{i^{k}_{j,r}}{m}< \frac{i}{m}\right\}z^{k}_{j}\left(i^{k}_{j,r}\right).
\ee
When $m$ tends to infinity
\[
i/m\to t,\quad i^{k}_{j,r}/m\to t^{k}_{j,r},\quad  z^{k}_{j}\left(i^{k}_{j,r}\right)\to z^{k}_{j,r},\quad Z^l_{(q-N_{l-1}-1)m+i}\to Z_q(t).
\]
These together yield $\int_0^1[z+Z_q(t)]^2\d t$
as shown in Eqs.~(\ref{QNL}) and (\ref{Zqt}). For the $m\to\infty$ limit of (\ref{Zldot}) first we rewrite it as
\be
\overline{Z^l_{^\cdot}}
=\frac{1}{n_l}\sum_{q\in C_l}\left[-\sum_{j=1}^{q-1}\sum_{r=1}^{\alpha^q_j}\left(q-N_{l-1}-1+\frac{i^q_{j,r}}{m}\right)\,z^{q}_{j}(i^q_{j,r}) +\sum_{k=q+1}^{N}\sum_{r=1}^{\alpha^k_q}\left(q-N_{l-1}-1+\frac{i^k_{q,r}}{m}\right)\,z^{k}_{q}(i^k_{q,r})\right].
\ee
Interchanging the order of summations with respect to $q$ and $j$ and to $q$ and $k$ and letting $m$ go to infinity one obtains the second form of $\overline{Z^l_{^\cdot}}$, see Eq.~(\ref{intZqbis}). In particular,
\be
\int_0^1 Z_q(t)\d t=
-\sum_{j=1}^{q-1}\sum_{r=1}^{\alpha^q_j}\left(q-N_{l-1}-1+t^q_{j,r}\right)\,z^{q}_{j,r} +\sum_{k=q+1}^{N}\sum_{r=1}^{\alpha^k_q}\left(q-N_{l-1}-1+t^k_{q,r}\right)\,z^{k}_{q,r}.
\ee
The $m\to\infty$ limit of (\ref{Zldotsquare}) by using (\ref{zji}) is
\bea\label{avZl^2}
n_l\overline{\left(Z^l_{^\cdot}\right)^2}
&=&\sum_{k,k'\in C_l}\sum_{j=1}^{k-1}\sum_{j'=1}^{k'-1} \sum_{r=1}^{\alpha^k_j}\sum_{r'=1}^{\alpha^{k'}_{j'}}A_{kk'}z^k_{j,r}\cdot z^{k'}_{j',r'}
-2\sum_{k,j'\in C_l}\sum_{j=1}^{k-1}\sum_{k'=j'+1}^{N}\sum_{r=1}^{\alpha^k_j}\sum_{r'=1}^{\alpha^{k'}_{j'}}A_{kj'} z^k_{j,r}\cdot z^{k'}_{j',r'}\nonumber\\
&+&\sum_{j,j'\in C_l}\sum_{k=j+1}^{N}\sum_{k'=j'+1}^{N}\sum_{r=1}^{\alpha^k_j}\sum_{r'=1}^{\alpha^{k'}_{j'}} A_{jj'}z^k_{j,r}\cdot z^{k'}_{j',r'}
\eea
where
\bea\label{Ak-first}
A_{kk'}=\min\left\{k+t^k_{j,r},\,k'+t^{k'}_{j',r'}\right\}-N_{l-1}-1,\nonumber\\
A_{kj'}=\min\left\{k+t^k_{j,r},\,j'+t^{k'}_{j',r'}\right\}-N_{l-1}-1,\nonumber\\
A_{jj'}=\min\left\{j+t^k_{j,r},\,j'+t^{k'}_{j',r'}\right\}-N_{l-1}-1.
\eea
To $A$ we added as subscripts only the two integers that run from $N_{l-1}+1$ to $N_l$; this is sufficient to distinguish the three cases. In the middle term the symmetry between $(j,k,r)$ and $(j',k',r')$ has been broken by contracting two equal sums.
Equation~(\ref{avZl^2}) is in pair with the second form of  $\overline{Z^l_{^\cdot}}$ in Eq.~(\ref{intZqbis}). This latter can be used to compute
the difference $\overline{\left(Z^l_{^\cdot}\right)^2}-\overline{Z^l_{^\cdot}}^2$. Starting with (\ref{avZl^2}),
the replacements
\bea
A_{kk'} \to A_{kk'}- \frac{1}{n_l}(k+t^k_{j,r}-N_{l-1}-1)(k'+t^{k'}_{j',r'}-N_{l-1}-1)\nonumber\\
A_{kj'} \to A_{kj'}-\frac{1}{n_l}(k+t^k_{j,r}-N_{l-1}-1)(j'+t^{k'}_{j',r'}-N_{l-1}-1)\nonumber\\
A_{jj'} \to A_{jj'}-\frac{1}{n_l}(j+t^k_{j,r}-N_{l-1}-1)(j'+t^{k'}_{j',r'}-N_{l-1}-1)
\eea
give $n_l[\overline{\left(Z^l_{^\cdot}\right)^2}-\overline{Z^l_{^\cdot}}^2]$. Clearly, all the three differences are nonnegative. In some cases another form of $\overline{\left(Z^l_{^\cdot}\right)^2}$, the analogue of the first form of $\overline{Z^l_{^\cdot}}$, may be useful. It can be obtained from (\ref{avZl^2}) by cutting four sums into two: $\sum_{j=1}^{k-1}=\sum_{j=1}^{N_{l-1}}+\sum_{j=N_{l-1}+1}^{k-1}$, $\sum_{k=j+1}^{N}=\sum_{k=j+1}^{N_l}+\sum_{k=N_l+1}^{N}$,
and similar for the sums with respect to $j', k'$. This yields
\bea
n_l\overline{\left(Z^l_{^\cdot}\right)^2}
&=&
\left[\sum_{\{j<k\}\subset C_{l}}\ \sum_{\{j'<k'\}\subset C_l} \sum_{r=1}^{\alpha^k_j}\sum_{r'=1}^{\alpha^{k'}_{j'}}(A_{kj'k'}-A_{jj'k'})
\right.\nonumber\\
&+&\left.\sum_{j,j'=1}^{N_{l-1}}\sum_{k,k'\in C_l} \sum_{r=1}^{\alpha^k_j}\sum_{r'=1}^{\alpha^{k'}_{j'}}A_{kk'}
+\sum_{j,j'\in C_l}\sum_{k,k'=N_l+1}^N   \sum_{r=1}^{\alpha^k_j}\sum_{r'=1}^{\alpha^{k'}_{j'}}A_{jj'}\right.
\nonumber\\
&+&\left.2\sum_{j=1}^{N_{l-1}}\sum_{k\in C_l}\sum_{\{j'<k'\}\subset C_l} \sum_{r=1}^{\alpha^k_j}\sum_{r'=1}^{\alpha^{k'}_{j'}}A_{kj'k'}
 -2\sum_{j=1}^{N_{l-1}}\sum_{k,j'\in C_l}\sum_{k'=N_l+1}^N \sum_{r=1}^{\alpha^k_j}\sum_{r'=1}^{\alpha^{k'}_{j'}} A_{kj'}\right.
 \nonumber\\
&-&\left. 2\sum_{\{j<k\}\subset C_{l}}\sum_{j'\in C_l}\sum_{k'=N_l+1}^N \sum_{r=1}^{\alpha^k_j}\sum_{r'=1}^{\alpha^{k'}_{j'}}A_{j'jk}
\right]
z^k_{j,r}\cdot z^{k'}_{j',r'}.
\eea
All the sums extend to the scalar product. The explicit $l$-dependence drops from the differences
\bea
A_{kj'k'}=A_{kk'}-A_{kj'}&=&\left\{\begin{array}{lll}
k'-j'&\mbox{if}&k'< k\\
k'-j'+\min\{t^k_{j,r},t^{k'}_{j',r'}\}-t^{k'}_{j',r'}&\mbox{if}&k'=k\\
k-j'+t^k_{j,r}-t^{k'}_{j',r'}&\mbox{if}&j'< k<k'\\
t^k_{j,r}-\min\{t^k_{j,r},t^{k'}_{j',r'}\}&\mbox{if}&j'=k\\
0&\mbox{if}&k<j'
\end{array}\right.
\nonumber\\
A_{jj'k'}=A_{jk'}-A_{jj'}&=&\left\{\begin{array}{lll}
k'-j'&\mbox{if}&k'< j\\
k'-j'+\min\{t^k_{j,r},t^{k'}_{j',r'}\}-t^{k'}_{j',r'}&\mbox{if}&k'=j\\
j-j'+t^k_{j,r}-t^{k'}_{j',r'}&\mbox{if}&j'< j<k'\\
t^k_{j,r}-\min\{t^k_{j,r},t^{k'}_{j',r'}\}&\mbox{if}&j'=j\\
0&\mbox{if}&j<j'
\end{array}\right.
\nonumber\\
A_{j'jk}=A_{kj'}-A_{jj'}&=&\left\{\begin{array}{lll}
k-j&\mbox{if}&k< j'\\
k-j+\min\{t^k_{j,r},t^{k'}_{j',r'}\}-t^{k}_{j,r}&\mbox{if}&k=j'\\
j'-j+t^{k'}_{j',r'}-t^{k}_{j,r}&\mbox{if}&j< j'<k\\
t^{k'}_{j',r'}-\min\{t^k_{j,r},t^{k'}_{j',r'}\}&\mbox{if}&j=j'\\
0&\mbox{if}&j'<j.
\end{array}\right.
\eea
The reader can write down the explicit form of $A_{kj'k'}-A_{jj'k'}$ which falls into $5+5$ subcases ($j,k,r$ and $j',k',r'$ interchanged).

The last point to check is that the $m\to\infty$ limit could indeed be taken under the summation signs. The cycle-dependent part of the summand of $G_m\left[\{n_l\}_1^p\right]$,
\be
f_{\left[\{n_l\}_1^p\right]}\left(\{t^k_{j,,r}\}_{r=1}^{\alpha^k_j},\{z^k_{j,r}\}_{r=1}^{\alpha^k_j}\right)= \prod_{l=1}^p \delta_{Z^l_1,0}\exp\left\{-\frac{\pi n_l\lambda_\beta^2}{L^2}\left[\overline{\left(Z^l_{^\cdot}\right)^2}-\overline{Z^l_{^\cdot}}^2\right]\right\}\sum_{z\in\Zz^d} \exp\left\{-\frac{\pi n_l\lambda_\beta^2}{L^2}\left(z+\overline{Z^l_{^\cdot}}\right)^2\right\}
\ee
is bounded as
\be\label{fbound}
|f_{\left[\{n_l\}_1^p\right]}|\leq \prod_{l=1}^p \sum_{z\in\Zz^d} \exp\left\{-\frac{\pi n_l\lambda_\beta^2}{L^2}z^2\right\}.
\ee
In $G_m\left[\{n_l\}_1^p\right]$ for each pair $j<k$ we cut the sum $\sum_{\alpha^k_j=0}^m$ into two parts, as we did for $N=2$. Then, using (\ref{fbound}) the argument given for $N=2$ can be repeated.  This concludes the proof of the Lemma.

\newsec{Summary}

In this paper we were engaged in finding an alternative to the Feynman-Kac formula for applications in quantum statistical physics. Our starting idea was to make somehow the integrations with respect to  the spatial variables possible.
If the integrand can be Fourier-expanded, the spatial variables appear linearly in the argument of complex units, their multipliers can be collected and the integrations over the torus executed. By the procedure the stochastic integrals would also be done implicitly.

Fourier expansion can be performed only after a discrete-time approximation. To collect the Fourier variables that multiply a given spatial variable is then straightforward but cumbersome and plenty of pitfalls. The problem we faced was partly notational: we had to deal with a huge number of variables, and without finding a suitable notation it would have been impossible to see through the formulas. A further complication is that different parts of the work require different notations.

A difficulty arose from the linear relations among the Fourier variables. These arrived in two stages. When integrating over the torus, the summed-up multipliers of the spatial variables are forced to be zero. Thus, a certain number of them can be eliminated by expressing them as a linear combination of the others. The dilemma of the choice was resolved by the fact that the Fourier variables associated with the Wiener measures, that is, the kinetic energy are just in the right number. So we eliminated them, and what was left is a function solely of the variables associated with pairs of interacting particles.
All those belonging to two particles in different cycles must satisfy a set of coupled homogeneous linear equations. The way these equations are coupled is nontrivial and could be understood with the help of graph theory.

Finally, the return to continuous variables is perhaps the most delicate point of all the discretizing methods. In the present case also the return to continuous time was tricky. We found the good idea by a detailed treatment of the simplest system, that of two particles. The paper ends with some awkward formulas: they are necessary for applications, not that easy to compute without making errors, so they can be helpful for the potential user.

\vspace{10pt}\noindent
{\bf Acknowledgements.} I am indebted to G\'abor Oszl\'anyi for his numerical analysis of the spectrum of the matrix $A$ appearing in Proposition \ref{spec(A)}. A correspondence with L\'aszl\'o Lov\'asz about the graph problem is gratefully acknowledged.

\newsec*{Appendix. Merger graphs}

We start with a class of graphs slightly different from the one presented under Remark \ref{constraint} but interesting in its own right. A graph of $V$ vertices and $E$ edges in this class corresponds to a system of $V$ homogeneous linear equations for $E$ variables, each appearing with coefficient 1, that has a solution in which all the variables take a nonzero integer value.

\begin{definition}\label{merge-gen1}
1.  A (merger) generator is a circle of even length, two odd circles with a common vertex, or two odd circles joining through a vertex the opposite endpoints of a linear graph. 2. The generators are mergers;  merging two mergers through one or more vertices and/or along one or more edges provides a merger. A graph composed of two disconnected mergers is a merger. 3. Given a merger, a set of generators whose merging provides the graph is called a covering. A covering is minimal if each of its elements contributes to the merger with at least one edge not covered by the other generators. A max-min covering is a minimal covering that contains the largest number of generators.
\end{definition}

As an example, the Petersen graph is a merger obtained by merging five "washtub" hexagons. The external edges are covered by three, the middle ones by two, the internal edges by one of the hexagons. Bipartite graphs, suitable subgraphs of the triangular lattice, complete graphs of more than three vertices are also mergers. The definition extends to multigraphs whose construction then involves also two-circles. In general the max-min covering is not unique, but the number of its constituting generators is uniquely determined because of the maximal property.

\begin{proposition}\label{merger1}
A graph is a merger if and only if to every edge one can assign a nonzero number in such a way that at every vertex the sum of the numbers assigned to the incident edges is zero. The numbers as variables form a manifold whose dimension is greater than or equal to the number of generators in the max-min coverings.
\end{proposition}

\noindent{\em Proof.} To mark the edges of a generator one can use a single and only a single variable denoted by $x_i$ for the $i$th generator of a merger. To the edges of a circle of length $2n$ one assigns $x_i$ and $-x_i$ in alternation. The same can be done with two odd circles sharing a vertex. In the case of two odd circles linked by a linear graph one assigns $x_i$ to those two edges of one of the circles that join the vertex of degree 3, and $-x_i$ and $x_i$ in alternation to the remaining edges of the same circle.
One then continues by alternating $-2x_i$ and $2x_i$ on the edges of the linear graph until reaching the second circle whose edges can again be marked by $x_i$ and $-x_i$ in a proper alternation. When merging, the generators carry their numbers. The number on a multiply covered edge is the sum of the numbers of the covering edges while one keeps the original number for the singly covered edges. Changing some $x_{i}$ in case of an accidental cancellation the sums on the edges are nonzero and the constraint remains satisfied.
The number of free variables cannot be smaller than the number of elements in max-min coverings, but it can be larger.
By construction, if the free variables are chosen to be integers, the others also take integers values.

In the opposite direction the proof goes by noting first that a graph does not contain any merger generator as a subgraph if and only if each of its maximal connected components is without circles or contains a single circle of odd length. The edges of such graphs cannot be marked in the required manner because either they have a vertex of degree 1 or they are unions of disjoint odd circles. Let $\cal G$ be any graph with properly marked edges. Thus, it has subgraphs which are merger generators. We can proceed by successive demerging. Let us choose a generator $g$ in $\cal G$ and select one of its edges denoted by $e$. Let $x$ be the number assigned to $e$. Prepare an image $g'$ of $g$ outside $\cal G$ and assign $-x$ to the image $e'$ of $e$. This uniquely determines the numbering of the other edges of $g'$ in such a way that the constraint is satisfied. Add the number on every edge of $g'$ to the number on its pre-image in $g$. As a result, the new number on $e$ is zero. Dropping all the edges from $\cal G$ whose new number is zero
and dropping also the vertices that become isolated we obtain a new graph having at least one edge less than $\cal G$ while the total numbering still satisfies the constraint. In a finite number of steps we can empty $\cal G$ which is, therefore, a merger. $\Box$

Now we define the class of graphs that we need for this paper. The vertices of a graph in this class correspond to permutation cycles, its edges represent the nonzero $\alpha^k_j$ that allow the solution of all the equations $Z^l_1=0$, cf. (\ref{Z^l_1}), with each variable taking a nonzero integer vector value.
We shall refer to these graphs as mergers generated by circles. The forthcoming discussion is restricted to them.

\begin{definition}\label{merge-gen2}
1. A (merger) generator is a circle of any (even or odd) length $n\geq 2$ with $n$ different positive integers assigned to the vertices in an arbitrary order. 2. The generators are mergers. Merging two mergers through all their vertices that carry the same number and optionally along some of the edges whose endpoints are common in the two mergers provides a merger. A graph composed of two disconnected mergers with disjoint vertex-numbering is a merger. 3. Given a merger, a set of generators whose merging provides the graph is called a covering. A covering is minimal if each of its element contributes to the merger with at least one edge not covered by the other generators. A max-min covering is a minimal covering that contains the largest number of generators.
\end{definition}

\begin{proposition}\label{merger2}
A graph whose vertices $\{1,2,\dots\}$ carry different positive integers $l_1,l_2,\dots$ is a merger if and only if to every edge one can assign a nonzero vector in such a way that at any vertex $i$ the sum of the vectors on the incident edges $(i,j)$ taken with minus sign if $l_j<l_i$ and with plus sign if $l_j>l_i$, is zero.
If $N_I$ and $M$ denote the number of linearly independent vectors and the number of generators in the max-min coverings, respectively, then $N_I\geq M$.
\end{proposition}

\noindent
In Proposition \ref{K=V-1} we shall give the precise value of $N_I$.

\vspace{5pt}
\noindent{\em Proof.} First let us see how to assign a vector to the edges of a generator. Let $l_i$ be the number carried by the $i$th vertex of a $n$-circle, where $i=1,\dots,n$ label the clockwise consecutive vertices. Let $x_{12}, x_{23},\dots,x_{n-1,n}, x_{n1}$ denote the $n$ edge variables that must assume a suitable value. The equation to be solved at vertex $i$  ($n+1\equiv 1$) is one of
\[
(1)\ \ x_{i-1,i}+x_{i,i+1}=0\quad \mbox{if   $l_{i-1},l_{i+1}>l_i$},\quad (2)\ \  -x_{i-1,i}+x_{i,i+1}=0\quad \mbox{if   $l_{i-1}<l_i<l_{i+1}$},
\]
\[
(3)\ \ x_{i-1,i}-x_{i,i+1}=0\quad \mbox{if   $l_{i-1}>l_i>l_{i+1}$},\quad (4)\ \  -x_{i-1,i}-x_{i,i+1}=0\quad \mbox{if   $l_{i-1},l_{i+1}<l_i$}.
\]
It is seen that whatever be the choice of, say, $x_{12}$, the other variables must take the same value with plus or minus sign. So the solution, if any, is a one-dimensional manifold.
To be definite, let $l_1$ be the smallest number. Take an arbitrary nonzero vector $v$ and set $x_{12}=v$. The equation at vertex 1 is (1), it is solved with $x_{n1}=-v$. We must prove that going around the circle there is no "frustration", all the equations can be solved. Call $i$ a source if $l_i<l_{i-1},l_{i+1}$ and a sink if $l_i>l_{i-1},l_{i+1}$. It is helpful to imagine an arrow on every edge, pointing towards the larger-numbered vertex. The problem is soluble because the number of sinks equals the number of sources and there is at least one source. Passing a source $-v$ changes to $v$ while solving equation (1), stays $v$ until the next sink and solves equations of the type (2), passing the sink it changes to $-v$ while solving equation (4), stays $-v$ and solves equations of the type (3), and so on.
By merging the circles of a covering as written down in Proposition~\ref{merger1} provides the solution for the merger graph.

The proof in the opposite direction goes again by successive demerging as described above provided we can show that no other vertex-numbered graph than those defined as mergers can be edge-marked in the required manner. Suppose there is such a finite graph. It must contain at least one circle, since linear graphs, tree graphs obviously cannot satisfy the condition at vertices of degree 1. Demerging successively all the circles, what remains is nonempty and cannot be marked -- a contradiction.  $\Box$

As a matter of fact, the numbering can be dropped from the definition because the vertices of any merger of circular graphs can be labelled {\em a posteriori} and in an arbitrary order with different $l_1,l_2,\dots$, still a proper assignment of nonzero vectors to the edges is possible.

\vspace{10pt}\noindent
{\em Example 1.}  A multigraph of two vertices and $n>1$ edges is a merger of $n-1$ two-circles.  Let $x_1,\dots,x_n$ be the edge variables. The equations at the two vertices are $\pm(x_1+\cdots+x_n)=0$. Their general solution with nonzero vectors is $x_1=v_1$, $x_2=-v_1+v_2$,\dots, $x_{n-1}=-v_{n-2}+v_{n-1}$, $x_n=-v_{n-1}$, a $(n-1)$-dimensional manifold. On the other hand, for $n$ even the collection of every second 2-circles constitutes a max-min covering. For $n$ odd we need one more circle. So $N_I=n-1$ and $M=n/2$ or $(n+1)/2$.

\noindent
{\em Example 2.} Consider the complete 4-graph of vertices 1, 2, 3, 4 with $l_i=i$. A max-min covering is for example the three circles $(123)$, $(234)$ and $(134)$. If prior to merging
\begin{eqnarray*}
z^2_1=x,\quad z^3_1=-x,\quad z^3_2=x,\quad\mbox{for $(123)$}\\
z^3_2=y,\quad z^4_2=-y,\quad z^4_3=y\quad\mbox{for $(234)$}\\
z^3_1=z,\quad z^4_1=-z,\quad z^4_3=z\quad\mbox{for $(134)$}\\
\end{eqnarray*}
then after merging the edges of the tetrahedron will carry
\begin{equation*}
z^2_1=x,\quad z^3_1=-x+z,\quad z^3_2=x+y,\quad
z^4_1=-z,\quad z^4_2=-y,\quad z^4_3=y+z\\
\end{equation*}
which solve the four equations
\[
z^2_1+z^3_1+z^4_1=0,\quad -z^2_1+z^3_2+z^4_2=0,\quad -z^3_1-z^3_2+z^4_3=0,\quad -z^4_1-z^4_2-z^4_3=0.
\]
A covering which is minimal but not max-min is the two 4-circles $(1234)$, $(1243)$. Using them we would get only two independent variables. A covering which is not minimal is obtained by merging e.g. $(1234)$ to the above three triangles with a fourth variable $v$. This only changes $x$ to $x'=x+v$ and $z$ to $z'=z+v$ without increasing the number of free variables. Now $N_I=M$.

\noindent
{\em Example 3.} Consider a part of the  triangular lattice composed of four small triangles, three with top up and one in the middle with top down. Let the V=6 vertices be numbered from 1 to 6 along the big outer triangle. This is a merger graph in both senses, generated by $M= 3$ elements in a max-min covering: according to Definition~\ref{merge-gen1}, by the rhombi $(1246)$, $(2346)$ and $(2456)$, and according to Definition~\ref{merge-gen2}, by the triangles $(126)$, $(234)$ and $(456)$.
We focus on mergers of the second kind.
Let the E=9 edge variables form the vector $Z$ with  $Z=(z^2_1\,  z^3_2\,  z^4_3\,  z^5_4\,  z^6_5\,  z^6_1\,  z^4_2\,  z^6_2\,  z^6_4)^T$ and let $A$ be the $V\times E$ matrix of the coefficients in the system of equations $AZ=0$ ($A$ is the incidence matrix of the directed graph with arrows towards the larger-numbered vertices).
One can check that ${\rm rank} A=5$, so 5 variables take a unique value once for the $N_I=4$ free variables a nonzero value is chosen. Thus, $N_I>M$. This can also be seen by choosing first a value for the variables associated with the three generators, as it was done in the proof of the propositions above. For the triangles, let $z^2_1=z^6_2=-z^6_1=x$, $z^3_2=z^4_3=-z^4_2=y$, $z^5_4=z^6_5=-z^6_4=z$. We can add any $v$ to $z^4_2$ so that the new value is $-y+v$; if we change simultaneously $z^6_2$ into $x-v$ and $z^6_4$ into $-z+v$, the four vectors $x$, $-y+v$, $x-v$ and $-z+v$ are linearly independent and
the three equations at vertices 2, 4 and 6, namely
\[
-z^2_1+z^3_2+z^4_2+z^6_2=0,\qquad -z^4_2-z^4_3+z^5_4+z^6_4=0,\qquad -z^6_1-z^6_2-z^6_4-z^6_5=0
\]
remain satisfied. The equations at the 2-degree vertices 1, 3 and 5 are unchanged.

\begin{proposition}\label{K=V-1}
Let $\cal G$ be a connected merger of circles having $V$ vertices and $E$ edges. Let $K$ denote the maximal number of linearly independent equations among the $V$ equations for the edge variables. Then $K=V-1$ independently of $E$, so the number of free variables is $N_I=E-V+1$. Moreover, any $V-1$ equations are linearly independent.
\end{proposition}

\noindent{\em Proof.} We proceed by induction according to the number of vertices. For $V=2$ the claim holds true, see Example 1. Suppose that it holds for $V=n$ and let $A_n$ denote the matrix of the coefficients in the $n$ equations. By the induction hypothesis $K\equiv{\rm rank} A_n=n-1$. If the vertex $n+1$ is coupled to any vertex $i\leq n$ with at most a single edge then
\[
A_{n+1}=\begin{pmatrix}
A_n & {\rm diag}\left(b_1^{n+1},\ldots,b_n^{n+1}\right)\\
\phantom{.}&\phantom{.}\\
0      & -b_1^{n+1}\ \ldots\ -b_n^{n+1}
\end{pmatrix}
\]
where
\[
b_i^{n+1}=\left\{\begin{array}{ll}
1 & \mbox{if there is an edge between vertices $i$ and $n+1$}\\
0 & \mbox{otherwise,}
\end{array}
\right.
\]
and at least two of them is 1. In the general case when vertex $n+1$ is coupled to vertex $i$ with $r_i$ edges, the diagonal matrix is replaced by a block-diagonal one where the $i$th block is
\[
B_i=(b^{n+1}_{i,1}\ldots b^{n+1}_{i,r_i}), \qquad i=1,\dots,n.
\]
At the same time in the last line we have $-b^{n+1}_{1,1}\ \ldots\ -b^{n+1}_{n,r_n}$ with at least two nonzero elements.
Now in $A_{n+1}$ the first $n$ lines are linearly independent, and replacing any one with the last line the $n$ lines are still linearly independent. On the other hand, the sum of the $n+1$ lines gives zero. So ${\rm rank}A_{n+1}=n$. $\quad\Box$

\vspace{5pt}
\noindent
{\em Example 4.}
For plane graphs and graphs consisting of the vertices and edges of convex polyhedra $N_I$ is related to the Euler characteristic $V+F-E=2$. Comparison with it shows that $N_I=F-1$ where $F$ is the number of faces (the exterior face included for plane graphs). Example 2 is a tetrahedron with $N_I=F-1=3=M$. In the case of Example 3 the general solution can also be obtained by choosing freely a variable for all the four triangles and then merge them.
The procedure applies to any connected plane graph: demerge it into faces, assign to the edges of the circle around each interior face a value as was shown in the proof of Proposition~\ref{merger2}, and merge the faces. This is an alternative way to solve the associated system of equations.
For edge-connected plane graphs $M<N_I$ because to cover the edges joining a vertex of even degree greater than 2 half of the faces incident on that vertex is enough. For the square lattice $M/N_I\to 1/2$. If the degree is odd, more faces are necessary. For the honeycomb lattice $M/N_I\to 3/4$.

\noindent
{\em Example 5.} There is another set of circle-generated merger graphs for which the general solution of the system of equations is easily obtained: the complete $n$-graphs. In this case $N_I=n(n-1)/2-n+1=(n-1)(n-2)/2$, the number of edges of the complete $(n-1)$-graph. The variables of this latter, $z^k_j$ for $1\leq j<k\leq n-1$ are free; then with
\[
z^n_l=\sum_{j=1}^{l-1}z^l_j-\sum_{k=l+1}^{n-1}z^k_l\qquad l=1,\dots,n-1
\]
all the equations will be satisfied. (For the complete 4-graph this offers the alternative to choose first $z^2_1=x$, $z^3_2=y$, $z^3_1=z$ and conclude with $z^4_1=-x-z$, $z^4_2=x-y$, $z^4_3=y+z$.) Note that for the complete $n$-graphs $N_I=M$: the triangles $(j,k,n)$ with $j<k<n$ form a max-min covering and $M=(n-1)(n-2)/2$ indeed. The edges $(j,k)$ are covered by a single triangle and each edge $(l,n)$ is covered by $n-2$ triangles.

In general, finding $N_I$ edges that can carry free variables or, equivalently, a unimodular $(V-1)\times(V-1)$ submatrix of the incidence matrix (whose columns belong to a possible set of dependent variables) may be difficult. Instead, one can easily find $N_I$ circles that constitute a covering, and assign a free variable to each circle. The variable is then equipped with edge-dependent signs on the edges of the circle. Finally, on every edge we obtain the sum of the signed circle-variables belonging to the circles incident on that edge, and the linear equations at each vertex will be satisfied with nonzero integer vectors.

\newpage
\noindent{\Large\bf References}
\begin{enumerate}

\item[{[F1]}] Feynman R. P.: {\em Space-time approach to non-relativistic quantum mechanics.} Rev. Mod. Phys. {\bf 20}, 367-387 (1948).
\item[{[F2]}] Feynman R. P.: {\em Atomic theory of the $\lambda$ transition in helium.} Phys. Rev. {\bf 91}, 1291-1301 (1953).
\item[{[G1]}] Ginibre J.: {\em Some applications of functional integration in Statistical Mechanics.} In: {\em Statistical Mechanics and Quantum Field Theory}, eds. C. De Witt and R. Stora. Gordon and Breach, New York (1971).
\item[{[G2]}] Ginibre J.: {\em Reduced density matrices of quantum gases. I. Limit of infinite volume.} J. Math. Phys. {\bf 6}, 238-251 (1965).
\item[{[G3]}] Ginibre J.: {\em Reduced density matrices of quantum gases. II. Cluster property.} J. Math. Phys. {\bf 6}, 252-262 (1965).
\item[{[G4]}] Ginibre J.: {\em Reduced density matrices of quantum gases. III. Hard-core potentials.} J. Math. Phys. {\bf 6}, 1432-1446 (1965).
\item[{[K1]}] Kac M.: {\em On distributions of certain Wiener functionals.} Trans. Amer. Math. Soc. {\bf 65}, 1-13 (1949).
\item[{[K2]}] Kac M.: {\em On some connections between probability theory and differential and integral equations.} In: Proceedings of the Second Berkeley Symposium on Probability and Statistics, J. Neyman ed., Berkeley, University of California Press (1951).
\item[{[LHB]}] L\H orinczi J., Hiroshima F., Betz V.: {\em Feynman-Kac-Type Theorems and Gibbs Measures on Path Space.} De Gruyter, Berlin/Boston (2011).
\end{enumerate}

\end{document}